\documentclass[journal]{IEEEtran}
\ifCLASSINFOpdf
\else
\fi
\usepackage{cite}
\usepackage{amsmath}
\usepackage[pdftex]{graphicx} 
\usepackage{subfig}
\usepackage{array} 

\usepackage{booktabs} 
\usepackage{amssymb}
\usepackage{mathtools}
\usepackage{cancel} 
\usepackage{bm} 
\usepackage{esint} 
\usepackage{amsthm}
\usepackage{xfrac}
\usepackage{tikz}
\usepackage{tikz-3dplot}
\usetikzlibrary{backgrounds,3d,shadings,shapes.misc,decorations.pathmorphing,shapes,calc,fadings,shadows.blur}

\usepackage{refcount}
\usepackage{siunitx}

\usepackage{amssymb,amsmath}

\usepackage{bm}
\newcommand{\bs}[1]{\bm{\mathrm{#1}}}

\newcommand{\vect}[1]{\vec{#1}}
\newcommand{\nhat}{\ensuremath{\hat{n}}}

\newcommand{\vecprime}[1]{\vect{#1}^{\,\prime}}

\renewcommand{\r}{\left(\vect{r}\right)}
\newcommand{\rp}{\left(\vecprime{r}\right)}
\newcommand{\rrp}{\left(\vect{r}, \vecprime{r}\right)}

\newcommand{\rrpki}[1][]{\left(k_i, \vect{r}, \vecprime{r}\right)}

\newcommand{\rin}[1]{\left(\vect{r}\right)\big\rvert_{#1}}
\newcommand{\norin}[1]{\big\rvert_{#1}}
\newcommand{\matr}[1]{\bs{#1}}

\newcommand{\abs}[1]{\left \lvert #1 \right\rvert }

\newcommand{\pvint}{\dashint}

\newcommand{\junk}[1] {}

\def\Xint#1{\mathchoice
{\XXint\displaystyle\textstyle{#1}}%
{\XXint\textstyle\scriptstyle{#1}}%
{\XXint\scriptstyle\scriptscriptstyle{#1}}%
{\XXint\scriptscriptstyle\scriptscriptstyle{#1}}%
\!\int}
\def\XXint#1#2#3{{\setbox0=\hbox{$#1{#2#3}{\int}$}
\vcenter{\hbox{$#2#3$}}\kern-.5\wd0}}

\def\dashint{\Xint-}

\newcommand*\widebar[1]{%
  \hbox{%
    \vbox{%
      \hrule height 0.5pt 
      \kern0.3ex
      \hbox{%
        \kern-0.05em
        \ensuremath{#1}%
        \kern-0.05em
      }%
    }%
  }%
} 

\newcommand{\mathsout}[1]
{\bgroup\mathchoice
	{\sbox0{$\displaystyle{#1}$}%
		\usebox0\hspace{-\wd0}%
		\rule[0.5\ht0-0.2\dp0-.5pt]{.7\wd0}{0.8pt}\hspace{.3\wd0}}%
	{\sbox0{$\textstyle{#1}$}%
		\usebox0\hspace{-\wd0}%
		\rule[0.5\ht0-0.2\dp0-.5pt]{.7\wd0}{0.8pt}\hspace{.3\wd0}}%
	{\sbox0{$\scriptstyle{#1}$}%
		\usebox0\hspace{-\wd0}%
		\rule[0.5\ht0-0.2\dp0-.5pt]{.7\wd0}{0.8pt}\hspace{.3\wd0}}%
	{\sbox0{$\scriptscriptstyle{#1}$}%
		\usebox0\hspace{-\wd0}%
		\rule[0.5\ht0-0.2\dp0-.5pt]{.7\wd0}{0.8pt}\hspace{.3\wd0}}%
	\egroup}

\renewcommand{\epsilon}{\varepsilon}

\newcommand{\opL}[1][]{\ensuremath{\mathcal{L}_{#1}}} 
\newcommand{\opM}[1][]{\ensuremath{\mathcal{M}_{#1}}} 

\newcommand{\opMpv}[1][]{\mathsout{\opM[#1]}} 

\newcommand{\Lmat}[1][]{{\matr{L}_{#1}}}

\newcommand{\Mmat}[1][]{{\matr{M}_{#1}}}

\newcommand{\Mpvmat}[1][]{{\mathsout{\matr{M}}_{#1}}}

\renewcommand{\tt}{{(\mathrm{tt})}}

\newcommand{\figref}[1]{Fig.~\ref{#1}}
\newcommand{\secref}[1]{Section~\ref{#1}}

\usepackage{tcolorbox}

\usepackage{textpos}

\newcommand{\mySubtitle}[1]%
{%
	\begin{textblock}{14.0}(0.7, 2.9)
		\textbf{#1}%
	\end{textblock}%
}%

\newcommand{\green}[1]{\textcolor{green!0!black}{#1}}

\newcommand{\brown}[1]{\textcolor{black}{#1}}
\newcommand{\magenta}[1]{\textcolor{black}{#1}}

\newcommand{\Ecolor}[1]{{#1}}

\newcommand{\Jcolor}[1]{{#1}}

\newcommand{\Rhocolor}[1]{\brown{#1}}
\newcommand{\rhocol}[1]{\Ecolor{#1}}

\newcommand{\Er}[1][]{\Ecolor{\vect{E}_{#1}\r}}

\newcommand{\Enrin}[2][]{\Ecolor{\nhat \cdot \vect{E}_{#1}\rin{#2}}}

\newcommand{\Jr}[1][]{\ensuremath{\Jcolor{\vect{J}_{#1}\r}}}

\newcommand{\Jmat}[1][]{\ensuremath{\Jcolor{\matr{J}_{#1}}}}

\newcommand{\Grrp}[1][]{\ensuremath{\green{G_{#1}\rrp}}}

\newcommand{\Grrpki}[1][]{\ensuremath{\green{G\rrpki}}}

\newcommand{\gradrGrrp}[1][]{\ensuremath{\green{\nabla G_{#1}\rrp}}}

\newcommand{\rhor}[1][]{\rhocol{\rho_{#1}\r}}

\newcommand{\Phimat}[1][]{\magenta{\matr{\Phi}_{#1}}}

\newcommand{\phir}[1][]{\magenta{\phi_{#1}\r}}
\newcommand{\phirp}[1][]{\magenta{\phi_{#1}\rp}}

\newcommand{\gradrphir}[1][]{\magenta{\nabla\phi_{#1}\r}}
\newcommand{\gradrpphirp}[1][]{\magenta{\nabla\phi_{#1}\rp}}

\newcommand{\phirin}[2][]{\magenta{\phi_{#1}\rin{#2}}}

\newcommand{\gradrphirin}[2][]{\magenta{\nabla\phi_{#1}\rin{#2}}}

\newcommand{\Jrin}[2][]{\Jcolor{\vect{J}_{#1}\rin{#2}}}
\newcommand{\Jin}[2][]{\Jcolor{\vect{J}_{#1}\,\norin{#2}}}

\newcommand{\ndgPhimat}[1][]{\ensuremath{\Rhocolor{\matr{\Psi}_{#1}}}}

\makeatletter
\newlength\numerator@height
\newlength\frac@height
\newsavebox\numerator@box
\newsavebox\frac@box

\newcommand\dfracparens[3]{%
	\sbox{\numerator@box}{\ensuremath{#1}}%
	\sbox{\frac@box}{\ensuremath{\dfrac{#1}{#2}}}%
	\settoheight{\frac@height}{\usebox{\frac@box}}%
	\settoheight{\numerator@height}{\usebox{\numerator@box}}%
	\addtolength{\frac@height}{-\numerator@height}%
	\usebox{\frac@box}%
	\raisebox{\frac@height}{%
		\( \left( {#3} \right)
		\)}%
}
\makeatother

\usepackage{dsfont}
\newcommand{\1}[1][]{\mathds{1}_{#1}}

\begin{document}
%
%
\title{A Generalized Scalar Potential Integral Equation Formulation for the DC Analysis of Conductors}

%
%
%

\author{Shashwat~Sharma,~\IEEEmembership{Graduate Student Member,~IEEE,}
	and~Piero~Triverio,~\IEEEmembership{Senior Member,~IEEE}
	\thanks{S. Sharma was with the Edward S. Rogers Sr. Department of Electrical \& Computer Engineering, University of Toronto, Toronto, ON, M5S 3G4 Canada, e-mail: shash.sharma@mail.utoronto.ca.
		P. Triverio is with the Edward S. Rogers Sr. Department of Electrical \& Computer Engineering and with the Institute of Biomedical Engineering, University of Toronto, Toronto, ON, M5S 3G4 Canada, email: piero.triverio@utoronto.ca.}
	\thanks{This work was supported by Advanced Micro Devices, by the Natural Sciences and Engineering Research Council of Canada (Collaborative Research and Development Grants program), and by CMC Microsystems.}
	\thanks{Manuscript received $\ldots$; revised $\ldots$.}}

%
%

\markboth{IEEE Transactions on Antennas and Propagation}%
{Sharma \MakeLowercase{\textit{et al.}}: Generalized Scalar Potential Integral Equation Formulation}
%



\maketitle

\begin{abstract} 
	The electrostatic modeling of conductors is a fundamental challenge in various applications, including the prediction of parasitic effects in electrical interconnects, the design of biasing networks, and the modeling of biological, microelectromechanical, and sensing systems.
	The boundary element method (BEM) can be an effective simulation tool for these problems because it allows modeling three-dimensional objects with only a surface mesh.
	However, existing BEM formulations can be restrictive because they make assumptions specific to particular applications.
	For example, capacitance extraction formulations usually assume a constant electric scalar potential on the surface of each conductor and cannot be used to model a flowing current, nor to extract the resistance.
	When modeling steady currents, many existing techniques do not address mathematical challenges such as the null space associated with the operators representing the internal region of a conductor.
	We propose a more general BEM framework based on the electric scalar potential for modeling conductive objects in various scenarios in a unified manner.
  Restrictive application-specific assumptions are not made, and the aforementioned operator null space is handled in an intuitive and rigorous manner.
	Numerical examples drawn from diverse applications confirm the accuracy and generality of the proposed method.
\end{abstract}


%
\IEEEpeerreviewmaketitle


\section{Introduction}

\IEEEPARstart{T}{he} electrostatic solution of Maxwell's equations is a fundamental necessity in a variety of applications.
For example, the design and analysis of integrated circuit components requires extracting the capacitance and resistance of chip- and package-level interconnects~\cite{parasitic_review}.
An electrostatic analysis can also be essential when designing power delivery networks~\cite{pdn} or biasing networks for quantum computing hardware~\cite{qem_transmon}.

The boundary element method~(BEM) is an effective approach for simulating problems involving piecewise-homogeneous conductive objects in an infinite homogeneous or stratified surrounding medium~\cite{ChewWAF}.
This is often the case in realistic applications such as capacitance and resistance extraction problems, where the BEM has been extensively used~\cite{cap01,cap02,cap03,cap04,cap05,cap06,cap07,cap08,DCR01,DCR02,DCR03,book:DCRC}.
Electrostatic analysis with the BEM has also been proposed in the context of microelectromechanical systems~\cite{spiemems01,spiemems02,spiemems03,spieafm01}, and in the biological and chemical domains, such as brain tissue modeling~\cite{deflation04,deflation05} and the simulation of molecular interactions~\cite{spiemol01,spiemol02,spiemol03,spiemol04}.
In the BEM, the Poisson or Laplace equation for the electric scalar potential~$\phi$ is used to derive a scalar potential integral equation~(SPIE)~\cite{book:colton}, and the goal is to compute the potential and/or surface charge density distribution~($\rho_{\mathrm{s}}$) on each object.

Although the DC solution of Maxwell's equations with the BEM may appear to be a solved problem, existing BEM formulations have mostly been developed for specific scenarios, and make restrictive assumptions which prevent their generalization to a broader class of problems.
For example, when considering isolated conductors in space, as is the case in capacitance extraction problems, the underlying physics is usually presupposed in the formulation by assuming that the scalar potential is constant on each object, which may not be valid when the conductors are embedded in a lossy material.
A single Laplace equation is written for the region external to the objects, and the associated SPIE is solved for the static charge distribution~\cite{cap01,cap02,cap03,cap04,cap05,cap06,cap07,cap08}.
These methods are designed to model the case where no current flows through the conductors.
In contrast, some scenarios require modeling objects connected via terminals and allowing the flow of current, such as resistance extraction problems.
In these cases, an SPIE must be formulated also for the internal region of each object to take into account the spatial variation of~$\phi$.
However, in the literature, SPIE formulations which model a flowing current tend to make assumptions on the flow path, involve geometric simplifications, or do not consider arbitrary 3D geometries~\cite{DCR01,DCR02,DCR03,book:DCRC}.
These simplifications lead to improved computational efficiency, but can also restrict the applicability of the formulation.
Furthermore, existing formulations typically do not take into account the approximate null space of the internal SPIE associated with constant potentials~\cite{deflation01}.
For standard BEM discretization schemes~\cite{RWG}, this null space may cause numerical issues for highly conductive objects when the potential drop across the object is small.
Linear algebraic techniques such as deflation have been proposed to handle this null space~\cite{deflation01,deflation02,deflation03,deflation04,deflation05,deflation06}, but they may not be compatible with standard preconditioning techniques when an iterative method is used to solve the final system of equations.

In this article, we propose an SPIE for the electrostatic analysis of conductors and develop a generalized formulation suitable for any of the scenarios mentioned above.
Our method does not assume that the potential is constant in a conductor, but naturally yields a constant potential when there is no current flow, e.g., when modeling an isolated set of conductors, as in capacitance extraction problems.
For scenarios involving a flowing current, such as resistance extraction, prior knowledge of the flow path is not required, and geometric simplifications are not necessary.
Therefore, in contrast to existing methods, both capacitance and resistance extraction are possible with the same formulation.
Furthermore, we devise intuitive physics-based consistency conditions to handle the null space associated with the internal region to obtain an invertible system matrix, rather than using linear algebraic approaches such as deflation~\cite{deflation01,deflation02,deflation03,deflation04,deflation05,deflation06}.
This enables the use of a simple triangular mesh with a standard piecewise constant expansion for~$\phi$ and~$\rho_{\mathrm{s}}$, which is not possible if the null space is not handled correctly.
The proposed method also supports various types of excitation such as a known total charge on the object, an incident potential generated by a nearby charge distribution, a Th\'evenin equivalent circuit, an applied potential with respect to infinity, or a combination of the above.
Our method is simple to implement and can be applied for the electrostatic analysis of conductors in a unified manner for a variety of applications, including capacitance extraction~\cite{cap01,cap02,cap03,cap04,cap05,cap06,cap07,cap08}, resistance extraction~\cite{DCR01,DCR02,DCR03,book:DCRC}, atomic force microscopy~\cite{spieafm01,afm04,femafm01,afm05,afm02,afm03,afm01}, electrostatic sensing~\cite{sensing02,sensing03,sensing01,sensing04}, and electrostatic discharge analysis~\cite{esd02,esd01}.

The goal of this article is to provide a physical and mathematical description of the proposed novel formulation and to demonstrate its generality.
The inclusion of acceleration algorithms~\cite{FMAorig,pfftmain,AIMbles} to solve large problems is not considered here.
However, these algorithms can be incorporated into the proposed formulation because it makes use of standard matrix operators which arise in the boundary element method.
The proposed formulation is described in~\secref{sec:formulation}, and the various types of excitation supported are detailed in~\secref{sec:exc}.
A discretization scheme for the integral equations and consistency conditions is proposed in~\secref{sec:discr}, including a description of the choice of basis and testing functions which leads to a full-rank system of equations.
Numerical examples representing several applications are provided in \secref{sec:results} and demonstrate the accuracy and generality of the proposed approach.
Concluding remarks are provided in~\secref{sec:conclusion}.


\section{Formulation}\label{sec:formulation}

First, we consider a single conductor in free space and describe a technique to handle the null space associated with the SPIE for the region internal to the conductor.
The extension to an arbitrary number of objects is described in~\secref{sec:discr}.

\subsection{Scalar Potential Integral Equations}\label{sec:spies}

\begin{figure}[t]
	\centering
	\includegraphics[width=0.5\linewidth]{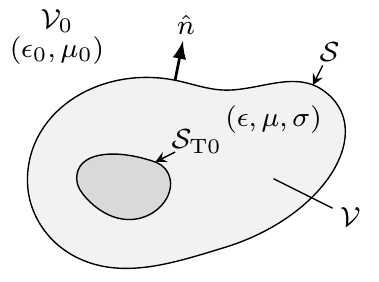}
	\caption{Geometry considered from \secref{sec:spies} to \secref{sec:appliedphi}: a single conductive object with a terminal~$\mathcal{S}_{\mathrm{T}0}$.}\label{fig:geom}
\end{figure}

Consider a conductive object occupying volume~$\mathcal{V}$ with surface~$\mathcal{S}$, outward unit normal vector~$\nhat$, permittivity~$\epsilon$, and conductivity~${\sigma>0}$, as shown in \figref{fig:geom}.
The object lies in free space denoted by~$\mathcal{V}_0$ with permittivity~$\epsilon_0$.
In the electrostatic case, the electric scalar potential~$\phir$ satisfies the Laplace equation for~${\vect{r}\in\mathcal{V}}$,
\begin{align}
	\nabla^2\phir = 0,\quad\left(\vect{r}\in\mathcal{V}\right).\label{eq:one:lapphiint}
\end{align}
Green's second identity~\cite{HansonYakovlev} can be used along with~\eqref{eq:one:lapphiint} to obtain an SPIE for the internal region~\cite{ChewWAF},
\begin{align}
	\opL\Bigl[\nhat'\cdot\gradrpphirp\Bigr] + \opM\Bigl[\phirp\Bigr] - \phir = 0,\label{eq:one:spieint0}
\end{align}
where~$\vect{r},\vecprime{r}\in\mathcal{S}^-$, with~$\mathcal{S}^-$ denoting the inner side of~$\mathcal{S}$.
In~\eqref{eq:one:spieint0}, the integral operators are defined as
\begin{align}
	\opL\Bigl[a\rp\Bigr] &= \int_{\mathcal{S}^-}d\mathcal{S}\,\Grrp\,a\rp,\label{eq:opL}\\
	\opM\Bigl[a\rp\Bigr] &= \int_{\mathcal{S}^-}d\mathcal{S}\,\nhat\cdot\gradrGrrp\,a\rp,\label{eq:opMpv}
\end{align}
where the static Green's function~$\Grrp$~\cite{ChewWAF} is
\begin{align}
	\Grrp = \frac{1}{4\pi \abs{\vect{r} - \vecprime{r}}}.\label{eq:hgf}
\end{align}

Similarly, starting from the Poisson equation for~${\vect{r}\in\mathcal{V}_0}$ with an impressed volume charge density~$\rhor[\mathrm{im}]$,
\begin{align}
	\nabla^2\phir = -\frac{\rhor[\mathrm{im}]}{\epsilon_0},\quad\left(\vect{r}\in\mathcal{V}_0\right),\label{eq:one:lapphiext}
\end{align}
an SPIE can be derived for the external region~\cite{ChewWAF},
\begin{align}
	\opL\Bigl[\nhat'\cdot\gradrpphirp\Bigr] + \opM\Bigl[\phirp\Bigr] + \phir = -\phir[\mathrm{im}],\label{eq:one:spieext0}
\end{align}
where~$\phir[\mathrm{im}]$ is the impressed potential generated by~$\rhor[\mathrm{im}]$, and~$\vect{r},\vecprime{r}\in\mathcal{S}^+$ with~$\mathcal{S}^+$ denoting the outer side of~$\mathcal{S}$.
In~\eqref{eq:one:spieext0}, the integrals associated with~$\opL$ and~$\opM$ are performed over~$\mathcal{S}^+$.

When~${\vect{r}\to\vecprime{r}}$,~$\Grrp$ becomes singular.
Consequently, the integral in~$\opM$ must be treated with the residue extraction procedure discussed in~\cite{ChewWAF,gibson},
\begin{align}
	\opM\Bigl[a\rp\Bigr] = \opMpv\Bigl[a\rp\Bigr] + \begin{cases}
		\frac{-1}{2}\,a\r,\quad\vect{r}\in\mathcal{S}^+,\\
		\frac{1}{2}\,a\r,\quad\vect{r}\in\mathcal{S}^-,
	\end{cases}
\end{align}
where the dash in~$\opMpv$ indicates that the associated integral is computed in the principal value sense~\cite{ChewWAF,gibson}.
The internal SPIE~\eqref{eq:one:spieint0} and the external SPIE~\eqref{eq:one:spieext0} then become
\begin{multline}
	\opL\Bigl[\nhat'\cdot\gradrpphirp\Bigr] + \opMpv\Bigl[\phirp\Bigr] - \frac{1}{2}\,\phir \\= 0,\quad\left(\vect{r},\vecprime{r}\in\mathcal{S}^-\right),\label{eq:one:spieint}
\end{multline}
and
\begin{multline}
	\opL\Bigl[\nhat'\cdot\gradrpphirp\Bigr] + \opMpv\Bigl[\phirp\Bigr] + \frac{1}{2}\,\phir \\= -\phir[\mathrm{im}],\quad\left(\vect{r},\vecprime{r}\in\mathcal{S}^+\right),\label{eq:one:spieext}
\end{multline}
respectively.

\subsection{Boundary Conditions}\label{sec:bc}

Next, boundary conditions are applied to relate the unknown quantities~$\phir$ and~$\nhat\cdot\gradrphir$ on~$\mathcal{S}^+$ and~$\mathcal{S}^-$.
The scalar potential is assumed continuous across~$\mathcal{S}$~\cite{PIE01,PIE02},
\begin{align}
	\phirin{\mathcal{S}^-} = \phirin{\mathcal{S}^+}.\label{eq:bcphi}
\end{align}
Since the electric field~${\Er=-\nabla\phir}$ inside a conductive object is zero in the static limit,
\begin{align}
	\nhat\cdot\gradrphirin{\mathcal{S}^-} = -\Enrin{\mathcal{S}^-} = 0.\label{eq:bcngphiin}
\end{align}
For~$\vect{r}\in\mathcal{S}^+$, we have
\begin{align}
	\nhat\cdot\gradrphirin{\mathcal{S}^+} = -\Enrin{\mathcal{S}^+} = -\frac{\rhor[\mathrm{s}]}{\epsilon_0},\label{eq:bcngphiout}
\end{align}
where~$\rho_{\mathrm{s}}$ denotes the surface charge density.

\subsection{Null Space of Operators Associated with~$\mathcal{V}$}\label{sec:nullsp}

Using the boundary condition~\eqref{eq:bcngphiin} in the internal SPIE~\eqref{eq:one:spieint}, we obtain
\begin{align}
	\opMpv\Bigl[\phirp\Bigr] - \frac{1}{2}\,\phir = 0.\label{eq:one:spieint2}
\end{align}
Consider the case where~$\phir$ equals a constant value~$\phi_{\mathrm{c}}$ on~$\mathcal{S}$, as is the case for a conductor at DC with no current flowing through.
Then,~\eqref{eq:one:spieint2} becomes
\begin{align}
	\phi_{\mathrm{c}}\,\opMpv\Bigl[1\Bigr] - \frac{1}{2}\,\phi_{\mathrm{c}} = 0,\label{eq:one:spieint3}
\end{align}
and one can show that~\cite{greenspdes}
\begin{align}
	\opMpv\Bigl[1\Bigr] = \pvint_{\mathcal{S}^-}d\mathcal{S}\,\nhat\cdot\gradrGrrp = \frac{1}{2}.\label{eq:opMconst}
\end{align}
Therefore, the left-hand side of~\eqref{eq:one:spieint3} is~$0$ regardless of the constant~$\phi_{\mathrm{c}}$, which implies that the operator in the internal SPIE~\eqref{eq:one:spieint2} has a null space corresponding to constant values of~$\phir$.
In other words, any constant~$\phi_{\mathrm{c}}$ satisfies~\eqref{eq:one:spieint2}, which is consistent with the physical intuition that the scalar potential of an isolated conductor at DC is unique up to a constant, and depends on the boundary conditions and on the choice of reference~\cite{PIE04,deflation01,deflation02}.
Upon the discretization of~\eqref{eq:one:spieint2} in a finite precision context, this null space can lead to ill-conditioned matrices and inaccurate results~\cite{deflation02}.
Weakly-varying potentials may also satisfy the discrete version of~\eqref{eq:one:spieint3} within numerical precision and hence may fall into the aforementioned null space, which may cause numerical issues even in the case of a steady current flow.
Conventional BEM-based capacitance solvers circumvent this issue by assuming a constant potential on~$\mathcal{S}$ and solving only the external SPIE~\eqref{eq:one:spieext} for~$\nhat\cdot\gradrphirin{\mathcal{S}^+}$~\cite{cap01,cap02,cap03,cap04,cap05,cap06,cap07,cap08}.
Existing resistance extraction methods do not address this null space issue~\cite{DCR01,DCR02,DCR03,book:DCRC}; instead, they seem to rely on linear basis functions for expanding~$\phir$ with improved accuracy, to avoid the numerical issues for weakly-varying potentials described above.
Linear algebraic approaches such as deflation have been proposed to handle this null space for biological applications~\cite{deflation01,deflation02,deflation03,deflation04,deflation05,deflation06}.
These methods manipulate the eigenvalue spectrum associated with the internal SPIE to allow solving the final system of equations with an iterative solver.
However, this approach may not be compatible with standard preconditioning techniques.
In the following section, we propose an intuitive physics-based approach to treat the null space of the internal SPIE~\eqref{eq:one:spieint2}, which offers physical insight and yields an invertible system of equations in all the scenarios considered here, including capacitance and resistance extraction.

\subsection{Extraction of the Average Potential}\label{sec:meanphi}

Rather than taking~$\phir$ as an unknown, we define a reduced potential~\cite{MTTVPIE}
\begin{align}
	\phir[\mathrm{r}] = \phir - \phi_{\mathrm{a}},\label{eq:phir}
\end{align}
where~$\phi_{\mathrm{a}}$ is the average potential on~$\mathcal{S}$,
\begin{align}
	\phi_{\mathrm{a}} = \frac{\int_{\mathcal{S}}d\mathcal{S}\,\phir}{\int_{\mathcal{S}}d\mathcal{S}} = \frac{1}{A}\int_{\mathcal{S}}d\mathcal{S}\,\phir,\label{eq:phiavg}
\end{align}
where~$A$ is the total area of~$\mathcal{S}$.
Therefore, the reduced potential~$\phir[\mathrm{r}]$ has zero mean,
\begin{align}
	\frac{1}{A}\int_{\mathcal{S}}d\mathcal{S}\,\phir[\mathrm{r}] = 0,\label{eq:phirem}
\end{align}
and captures only the spatial variations of~$\phir$ on~$\mathcal{S}$.
For an isolated conductor, we expect~${\phir[\mathrm{r}]=0}$, but this would no longer be true when the conductor is connected to a closed circuit and a steady current flows through it.
In order to handle both cases, we do not impose restrictions on~$\phir[\mathrm{r}]$; instead, we take both~$\phir[\mathrm{r}]$ and~$\phi_{\mathrm{a}}$ as separate unknown quantities.
The former allows modeling the spatial variation of~$\phi$ associated to a flowing current, while the latter is related to the fact that as charge accumulates on the object, its average potential with respect to the reference will increase.

Using~\eqref{eq:phir} in the internal SPIE~\eqref{eq:one:spieint2} to replace~$\phir$,
\begin{multline}
	\opMpv\Bigl[\phirp[\mathrm{r}]\Bigr] - \frac{1}{2}\,\phir[\mathrm{r}] + \opMpv\Bigl[\phi_{\mathrm{a}}\Bigr] \\- \frac{1}{2}\,\phi_{\mathrm{a}} = 0,\quad\left(\vect{r},\vecprime{r}\in\mathcal{S}^-\right),\label{eq:one:spieint4}
\end{multline}
where the linearity of~$\opMpv$ was used.
Since $\phi_{\mathrm{a}}$ is constant, we can use~\eqref{eq:opMconst} in~\eqref{eq:one:spieint4} to obtain,
\begin{align}
	\opMpv\Bigl[\phirp[\mathrm{r}]\Bigr] - \frac{1}{2}\,\phir[\mathrm{r}] = 0,\quad\left(\vect{r},\vecprime{r}\in\mathcal{S}^-\right).\label{eq:one:spieint5}
\end{align}
Since the average potential~$\phi_{\mathrm{a}}$ was extracted from~$\phir$, the reduced potential~$\phir[\mathrm{r}]$ represents \emph{only} the spatial variation of~$\phir$.
Therefore, a unique solution can be obtained for~$\phir[\mathrm{r}]$ regardless of the choice of reference for~$\phi$.
This is the key to avoiding the null space of the internal SPIE.
This concept was also exploited in the linear algebraic approach taken in some existing works~\cite{deflation01,deflation02,deflation03,deflation04,deflation05,deflation06}.

Similarly, using~\eqref{eq:phir} in the external SPIE~\eqref{eq:one:spieext} to replace~$\phir$,
\begin{multline}
	\opL\Bigl[\nhat'\cdot\gradrpphirp[0]\Bigr] + \opMpv\Bigl[\phirp[\mathrm{r}]\Bigr] + \frac{1}{2}\,\phir[\mathrm{r}] + \opMpv\Bigl[\phi_\mathrm{a}\Bigr] + \frac{1}{2}\,\phi_\mathrm{a} \\= -\phir[\mathrm{im}],\quad\left(\vect{r},\vecprime{r}\in\mathcal{S}^+\right).\label{eq:one:spieext2}
\end{multline}
Using~\eqref{eq:opMconst} in~\eqref{eq:one:spieext2} then gives
\begin{multline}
	\opL\Bigl[\nhat'\cdot\gradrpphirp[0]\Bigr] + \opMpv\Bigl[\phirp[\mathrm{r}]\Bigr] + \frac{1}{2}\,\phir[\mathrm{r}] + \phi_\mathrm{a} \\= -\phir[\mathrm{im}],\quad\left(\vect{r},\vecprime{r}\in\mathcal{S}^+\right).\label{eq:one:spieext3}
\end{multline}
The proposed formulation supports a variety of excitations, including a known total charge, an applied potential, or a circuit excitation through a set of ports, as described below.

%

\section{Types of Excitation}\label{sec:exc}

\subsection{Total Charge Specification}\label{sec:totQ}

Due to~\eqref{eq:one:spieint3} and~\eqref{eq:opMconst}, the discretized operator~${\opMpv - \sfrac{1}{2}}$ is rank deficient.
Therefore, uniquely determining the constant~$\phi_\mathrm{a}$ requires an additional equation per isolated object.
The additional equations are provided via the excitation.
For example, if the total charge~$Q$ on~$\mathcal{S}^+$ is known, it can be related to the unknown quantity~$\nhat\cdot\gradrphir$ as
\begin{align}
	Q = -\epsilon_0\int_{\mathcal{S}^+}d\mathcal{S}\,\nhat\cdot\gradrphir.\label{eq:Qdef}
\end{align}
Therefore, we can take~\eqref{eq:Qdef} as an additional equation in the system.

\subsection{Applied Potential}\label{sec:appliedphi}

In some applications, one may need to set a conductor at a fixed known potential with respect to infinity by attaching the object to a battery via a terminal.
A typical example is the capacitance extraction problem.
In our formulation, this can be accomplished by defining a small portion of~$\mathcal{S}$ as a terminal area denoted by~${\mathcal{S}_{\mathrm{T}0}\in\mathcal{S}}$ (\figref{fig:geom}), and setting the potential~$\phirin{\mathcal{S}_{\mathrm{T}0}}$ to a known value,
\begin{align}
	\phirin{\mathcal{S}_{\mathrm{T}0}} = \phi_{0}\label{eq:phiatST}
\end{align}
which will be taken as an additional equation \emph{instead} of the total charge condition~\eqref{eq:Qdef}, to solve as part of the final system of equations as described in \secref{sec:finalsys}.
It is assumed that the area of~${\mathcal{S}_{\mathrm{T}0}}$ is small enough that~$\phi_{0}$ is constant over~${\mathcal{S}_{\mathrm{T}0}}$.

\subsection{Attached Circuit}\label{sec:circuit}

\begin{figure}[t]
	\centering
	\subfloat[][]{%
	\includegraphics[width=0.44\linewidth]{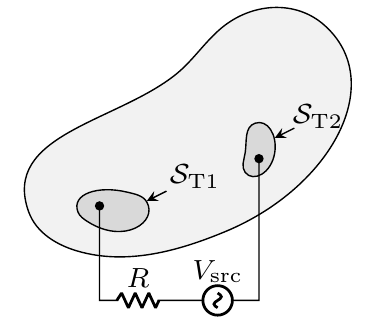}\label{fig:geom1}}
	\subfloat[][]{%
	\includegraphics[width=0.56\linewidth]{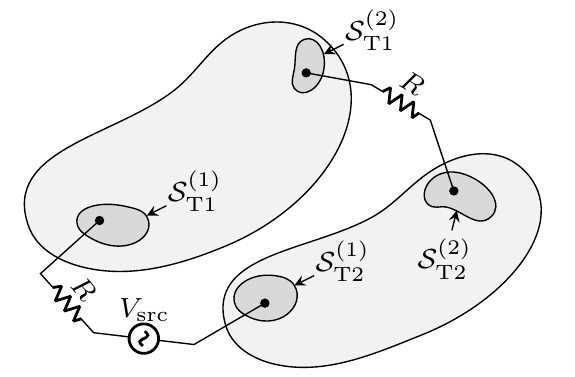}\label{fig:geom2}}
	\caption{Configurations considered in \secref{sec:circuit:oneobj} (a) and in \secref{sec:circuit:twoport} (b).}\label{fig:geom3}
\end{figure}

Finally, we consider the case of an attached circuit, which is assumed to be a Th\'evenin equivalent circuit for simplicity.
We consider two cases: the two terminals of the Th\'evenin equivalent circuit attached to a single conductor, where one behaves as a source and the other as a sink (\figref{fig:geom1}); and the case of a Th\'evenin equivalent circuit attached via ports to two conductors (\figref{fig:geom2}).
Considering these simple configurations is sufficient for generalizing the proposed method to arbitrary combinations of ports and conductors.

\subsubsection{Circuit attached to a single object}\label{sec:circuit:oneobj}

Consider the setup in \figref{fig:geom1}, where the two circuit terminals are denoted as~$\mathcal{S}_{\mathrm{T}1}$ and~$\mathcal{S}_{\mathrm{T}2}$.
Since a current density~$\Jr$ may flow through~$\mathcal{V}$, a new boundary condition must be derived for~$\nhat\cdot\gradrphir$ on~${\mathcal{S}^-_{\mathrm{T}t}}$ because~\eqref{eq:bcngphiin} is no longer valid.
Instead, we have
\begin{align}
	\nhat\cdot\gradrphirin{\mathcal{S}^-_{\mathrm{T}t}} = -\Enrin{\mathcal{S}^-_{\mathrm{T}t}} = -\frac{1}{\sigma}\,\nhat\cdot\Jrin{\mathcal{S}^-_{\mathrm{T}t}},\label{eq:ndgphiJ}
\end{align}
where~$t\in\{1,2\}$. 
Applying the boundary condition for the normal component of the volume current density for a lossy conductor at DC~\cite{rothwellEM} gives
\begin{align}
	\nhat\cdot\Jrin{\mathcal{S}^+_{\mathrm{T}t}} - \nhat\cdot\Jrin{\mathcal{S}^-_{\mathrm{T}t}} = 0.\label{eq:bcJn}
\end{align}
Assuming that~$\nhat\cdot\Jr$ is constant over~$\mathcal{S}^+_{\mathrm{T}t}$ and~$\mathcal{S}^-_{\mathrm{T}t}$, we can relate~${\nhat\cdot\Jin{\mathcal{S}^+_{\mathrm{T}t}}}$ to the current flowing through the Th\'evenin equivalent circuit~$I$ as
\begin{align}
	\nhat\cdot\Jin{\mathcal{S}^+_{\mathrm{T}t}} = \frac{\mp I}{A_{\mathrm{T}t}},\label{eq:IJT}
\end{align}
where~$A_{\mathrm{T}t}$ is the area of~$\mathcal{S}_{\mathrm{T}t}$, and the positive sign in front of~$I$ is taken when~$I$ flows out of the associated terminal (e.g.,~$\mathcal{S}_{\mathrm{T}2}$ in the case of \figref{fig:geom1}).
Using~\eqref{eq:bcJn} in~\eqref{eq:ndgphiJ} gives the desired boundary condition,
\begin{align}
	\nhat\cdot\gradrphirin{\mathcal{S}^-_{\mathrm{T}t}} = -\frac{1}{\sigma}\,J_{\mathrm{T}t} = \frac{\mp I}{A_{\mathrm{T}t}}.\label{eq:bcndgphiJ}
\end{align}
where we have defined
\begin{align}
	J_{\mathrm{T}t} = \nhat\cdot\Jin{\mathcal{S}^+_{\mathrm{T}t}}
\end{align}
for convenience.
The SPIE for the internal region~\eqref{eq:one:spieint5} must then be modified as
\begin{align}
	\opMpv\Bigl[\phirp[\mathrm{r}]\Bigr] - \frac{1}{2}\,\phir[\mathrm{r}] - \frac{1}{\sigma}\,\opL\Bigl[J_{\mathrm{T}t}\Bigr]= 0,\quad\left(\vect{r},\vecprime{r}\in\mathcal{S}^-\right),\label{eq:one:spieint6}
\end{align}
where~$J_{\mathrm{T}t}\r$ is non-zero only at terminals.
Since~$J_{\mathrm{T}t}$ is an additional known, additional equations can be obtained for~$J_{\mathrm{T}t}$ by writing the Kirchoff voltage law~(KVL) for the circuit in terms of the volume current density associated with each terminal,
\begin{align}
	\phirin{\mathcal{S}^-_{\mathrm{T}1}} - \phirin{\mathcal{S}^-_{\mathrm{T}2}} + R\,J_{\mathrm{T}1}A_{\mathrm{T}1} &= V_{\mathrm{src}},\label{eq:kvl1}\\
	\phirin{\mathcal{S}^-_{\mathrm{T}1}} - \phirin{\mathcal{S}^-_{\mathrm{T}2}} + R\,J_{\mathrm{T}2}A_{\mathrm{T}2} &= V_{\mathrm{src}},\label{eq:kvl2}
\end{align}
where~$R$ and~$V_{\mathrm{src}}$ are the resistance and voltage associated with the Th\'evenin equivalent circuit.
Note that the total charge on the object must also be specified using~\eqref{eq:Qdef} to get a square system of equations at the end.

\subsubsection{Two-port network}\label{sec:circuit:twoport}

Next, consider the two-port setup in \figref{fig:geom2}, involving two objects~$\mathcal{S}_1$ and~$\mathcal{S}_2$, where~${\mathcal{S}=\mathcal{S}_1\bigcup\mathcal{S}_2}$.
The two terminals associated with port~$i$ are denoted as~$\mathcal{S}_{\mathrm{T}1}^{(i)}$ and~$\mathcal{S}_{\mathrm{T}2}^{(i)}$, and the corresponding port current is~$I^{(i)}$.
The KVL can be written again for each port in terms of the volume current density associated with each terminal,
\begin{align}
	\phirin{\mathcal{S}^-_{\mathrm{T}1}} - \phirin{\mathcal{S}^-_{\mathrm{T}2}} + R\,J_{\mathrm{T}1}^{(i)}A_{\mathrm{T}1}^{(i)} = V_{\mathrm{src}}^{(i)},\label{eq:kvlt1}\\
	\phirin{\mathcal{S}^-_{\mathrm{T}1}} - \phirin{\mathcal{S}^-_{\mathrm{T}2}} + R\,J_{\mathrm{T}2}^{(i)}A_{\mathrm{T}2}^{(i)} = V_{\mathrm{src}}^{(i)},\label{eq:kvlt2}
\end{align}
where~${V_{\mathrm{src}}^{(2)}=0}$ for the setup in \figref{fig:geom2},~$J_{\mathrm{T}t}^{(i)}$ is the normal component of the volume current density associated with terminal~$t$ of port~$i$, and~$A_{\mathrm{T}t}^{(i)}$ is the corresponding surface area.
Note that the resistance~$R$ need not be the same for each port, but is assumed so here to simplify the notation.

Since there are two objects, an average surface potential is defined for each object~$q$ as
\begin{align}
	\phi_{\mathrm{a}}^{(q)} = \frac{\int_{\mathcal{S}_q}d\mathcal{S}\,\phir}{\int_{\mathcal{S}_q}d\mathcal{S}} = \frac{1}{A_q}\int_{\mathcal{S}_q}d\mathcal{S}\,\phir,\label{eq:phiavgq}
\end{align}
where~$\mathcal{S}_q$ denotes the surface of object~$q$.
Additional equations are needed to ensure that the total charge is specified and the current continuity condition is satisfied.
One equation can be obtained by specifying the total charge~$Q$ on the two connected objects,
\begin{align}
	Q &= -\epsilon_0\int_{\mathcal{S}^+}d\mathcal{S}\,\nhat\cdot\gradrphir\label{eq:Qdef2}\\
	&= -\epsilon_0\left[\int_{\mathcal{S}_1^+}d\mathcal{S}\,\nhat\cdot\gradrphir + \int_{\mathcal{S}_2^+}d\mathcal{S}\,\nhat\cdot\gradrphir\right].\label{eq:Qdef3}
\end{align}
Two additional equations are obtained by applying the Kirchoff current law~(KCL), treating each object as a circuit node.
Applying the KCL at each node yields
\begin{align}
	J_{\mathrm{T}1}^{(1)}A_{\mathrm{T}1}^{(1)} &= J_{\mathrm{T}1}^{(2)}A_{\mathrm{T}1}^{(2)},\label{eq:kcl1}\\
	J_{\mathrm{T}2}^{(1)}A_{\mathrm{T}2}^{(1)} &= J_{\mathrm{T}2}^{(2)}A_{\mathrm{T}2}^{(2)}.\label{eq:kcl2}
\end{align}
Note that only one of~\eqref{eq:kcl1} and~\eqref{eq:kcl2} is needed in the final system of equations, because the other is implied by the KVL equations~\eqref{eq:kvlt1} and~\eqref{eq:kvlt2}.
Generalization to the case of multiple objects and ports is discussed in the next section.

\section{Discretization and Final System of Equations}\label{sec:discr}

To obtain the final discretized system of equations, we consider a general setup involving~$N_{\mathrm{obj}}$ objects, where the surface of object~$q$ is denoted as~$\mathcal{S}_q$ and~${\mathcal{S}=\bigcup_q^{N_{\mathrm{obj}}}\mathcal{S}_q}$
Each object is either in isolation with a known total charge, connected to the terminal of one or more Th\'evenin equivalent circuit ports, or connected to a voltage supply with respect to infinity.
An example configuration involving each of these cases is depicted in the right and left panels of \figref{fig:geom3}, respectively.
A triangular mesh is generated for the surface of each object in the structure, and we assume that the mesh for object~$q$ contains~$N_{\mathrm{tri}}^{(q)}$ triangles.

\subsection{Choice of Basis Functions}\label{sec:discr:basis}

The quantities which will be taken as the final set of unknowns are~$\nhat\cdot\gradrphirin{\mathcal{S}^+}$,~$\phir[\mathrm{r}]$,~$\phi_{\mathrm{a}}$, and~$J_{\mathrm{T}t}$, and we next discuss the choice of basis functions for each of them.

Quantities~$\nhat\cdot\gradrphirin{\mathcal{S}^+}$ and~$J_{\mathrm{T}t}$ are expanded with area-normalized pulse functions~$h_n\r$ which are constant over the associated~$n\mathrm{th}$ mesh element.
The unknown coefficients associated with~$\nhat\cdot\gradrphirin{\mathcal{S}^+}$ and~$J_{\mathrm{T}t}$ are stored in the column vectors~$\ndgPhimat$ and~$\Jmat[\mathrm{T}]$.
For simplicity, we assume that~$\mathcal{S}_{\mathrm{T}t}$ is small and spans only one mesh triangle, so that there are as many terminal triangles as terminals.
In cases where a terminal must span multiple mesh elements, one may include an additional set of equations to enforce a constant potential over all the mesh elements associated with~$\mathcal{S}_{\mathrm{T}t}$~\cite{gope}.

To devise a strategy for discretizing the remainder term~$\phir[\mathrm{r}]$, we follow a procedure similar to the one proposed in~\cite{MTTVPIE}.
Observing from~\eqref{eq:phir} that~$\phirin[\mathrm{r}]{\mathcal{S}_q}$ has a zero average value over~$\mathcal{S}_q$, the surface of object~$q$, we seek a basis function for expanding~$\phirin[\mathrm{r}]{\mathcal{S}_q}$ that preserves this zero-mean property.
Suppose that column vector~${\Phimat_{\mathrm{r}}^{(q)}}$ contains the unknown coefficients associated with~$\phirin[\mathrm{r}]{\mathcal{S}_q}$, while~$\Phimat^{(q)}$ contains the coefficients associated with~$\phirin{\mathcal{S}_q}$, and both quantities are expressed in terms of pulse basis functions, and both column vectors have length~$N_{\mathrm{tri}}$.
Preserving the zero-mean property of~$\phirin[\mathrm{r}]{\mathcal{S}_q}$ requires that~${\Phimat_{\mathrm{r}}^{(q)}}$ belongs to a subspace of dimension~${(N_{\mathrm{tri}}^{(q)}-1)}$, which can be accomplished by seeking a basis~$\matr{D}_{\mathrm{r}}^{(q)}$ of dimension~${N_{\mathrm{tri}}^{(q)}-1}$ so that
\begin{align}
	\Phimat_{\mathrm{r}}^{(q)} = \matr{D}_{\mathrm{r}}^{(q)}\matr{v}_{\mathrm{r}}^{(q)},\label{eqd:phirt0}
\end{align}
where~${\matr{v}_{\mathrm{r}}^{(q)}\in\mathbb{C}^{(N_{\mathrm{tri}}^{(q)}-1)}}$.
As in~\cite{MTTVPIE},~${\matr{D}_{\mathrm{r}}^{(q)}\in\mathbb{R}^{N_{\mathrm{tri}}^{(q)}\times (N_{\mathrm{tri}}^{(q)}-1)}}$ is chosen as
\begin{align}
	\matr{D}_{\mathrm{r}}^{(q)} \triangleq \begin{bmatrix} \matr{I}_{\mathrm{r}} \\ -\left(\1^{(q)}\right)^T \end{bmatrix},\label{eqd:Drdef}
\end{align}
where~${\matr{I}_{\mathrm{r}}\in\mathbb{R}^{(N_{\mathrm{tri}}^{(q)}-1) \times (N_{\mathrm{tri}}^{(q)}-1)}}$ is the identity matrix and column vector~${\1^{(q)}\in\mathbb{R}^{N_{\mathrm{tri}}^{(i)}}}$ contains all ones.
As a result of this choice,~$\matr{v}_{\mathrm{r}}^{(q)}$ contains potentials with respect to the average surface potential on~$\mathcal{S}_q$, and is the quantity we will take as unknown in lieu of~$\Phimat_{\mathrm{r}}^{(q)}$.
Equation~\eqref{eq:phir} can then be written in the discrete domain for each object~$q$ as
\begin{align}
	\Phimat_{\mathrm{r}}^{(q)} = \matr{D}_{\mathrm{r}}^{(q)}\matr{v}_{\mathrm{r}}^{(q)} = \Phimat^{(q)} - \1^{(q)}\phi_\mathrm{a}^{(q)}.\label{eqd:phirt}
\end{align}


As in~\cite{MTTVPIE}, the vectors of scalar potential unknowns associated with each object can then be concatenated as
\begin{align}
	\matr{v}_{\mathrm{r}} =
	\begin{bmatrix}
		\matr{v}_{\mathrm{r}}^{(1)} \\ \vdots \\ \matr{v}_{\mathrm{r}}^{(N_{\mathrm{obj}})}
	\end{bmatrix},\quad
	\Phimat[\mathrm{a}] =
	\begin{bmatrix}
		\phi_{\mathrm{a}}^{(1)} \\ \vdots \\ \phi_{\mathrm{a}}^{(N_{\mathrm{obj}})}
	\end{bmatrix},\label{eqd:phiremat}
\end{align}
so that
\begin{align}
	\Phimat &= \matr{D}_{\mathrm{r}}\matr{v}_{\mathrm{r}} + \1\Phimat[\mathrm{a}],\label{eqd:phimat}
\end{align}
where
\begin{align}
	\matr{D}_{\mathrm{r}} &=
	\begin{bmatrix}
		\matr{D}_{\mathrm{r}}^{(1)} & \cdots & \matr{0} \\
		\vdots & \ddots & \vdots \\
		\matr{0} & \cdots & \matr{D}_{\mathrm{r}}^{(N_{\mathrm{obj}})}
	\end{bmatrix},\label{eqd:Dr}\\
	\1 &=
	\begin{bmatrix}
		\1^{(1)} & \cdots & \matr{0} \\
		\vdots & \ddots & \vdots \\
		\matr{0} & \cdots & \1^{(N_{\mathrm{obj}})}
	\end{bmatrix}.\label{eqd:1}
\end{align}

\subsection{Testing the Integral Equations}\label{sec:testing}

To obtain the final system of equations, the external and internal SPIEs for each object~\eqref{eq:one:spieext3} and~\eqref{eq:one:spieint6}, respectively, are tested with area-normalized pulse functions~$h_m\r$.
The discretized SPIE for the external region~\eqref{eq:one:spieext3} reads
\begin{align}
	\Lmat\ndgPhimat + \Mpvmat\matr{D}_{\mathrm{r}}\matr{v}_{\mathrm{r}} + \frac{1}{2}\matr{I}_A\matr{D}_{\mathrm{r}}\matr{v}_{\mathrm{r}} + \1\Phimat[\mathrm{a}] = -\Phimat[\mathrm{im}],\label{eqd:spieext}
\end{align}
where~$\Lmat$ and~$\Mpvmat$ are the discretized~$\opL$ and~$\opMpv$ operators, respectively.
Entries of column vector~$\Phimat[\mathrm{im}]$ are associated with~$\phir[\mathrm{im}]$, while~$\matr{I}_A$ is the identity matrix whose entries are scaled by the area of the triangle corresponding to each row.
For the internal region of each object~$q$, the SPIE~\eqref{eq:one:spieint6} in discrete form is
\begin{align}
	\Mpvmat^{(q)}\matr{D}_{\mathrm{r}}^{(q)}\matr{v}_{\mathrm{r}}^{(q)} - \frac{1}{2}\matr{I}_A^{(q)}\matr{D}_{\mathrm{r}}^{(q)}\matr{v}_{\mathrm{r}}^{(q)} - \frac{1}{\sigma^{(q)}}\,\Lmat^{(q)}\matr{D}_{\mathrm{T}}^{(q)}\Jmat_{\mathrm{T}}^{(q)} = \matr{0},\label{eqd:spieintq}
\end{align}
where~$\Lmat^{(q)}$ and~$\Mpvmat^{(q)}$ are the discretized~$\opL$ and~$\opMpv$ operators, respectively.
The superscript~$(q)$ on each term in~\eqref{eqd:spieintq} indicates that the corresponding term is associated with object~$q$.
The sparse incidence matrix~$\matr{D}_{\mathrm{T}}^{(q)}\in\mathbb{R}^{(N_{\mathrm{tri}}^{(q)} \times N_{\mathrm{term}}^{(q)})}$ maps from the~$N_{\mathrm{term}}^{(q)}$ terminal current densities associated with object~$q$ and stored in~$\Jmat_{\mathrm{T}}^{(q)}$, to the associated triangle in the mesh.
Note that the last term in~\eqref{eqd:spieintq} is zero for objects which are not connected to any terminal.

The SPIEs for the internal region of all objects can now be written together as
\begin{align}
	\Mpvmat_{\mathrm{in}}\matr{D}_{\mathrm{r}}\matr{v}_{\mathrm{r}} - \frac{1}{2}\matr{I}_A\matr{D}_{\mathrm{r}}\matr{v}_{\mathrm{r}} - \Lmat_{\mathrm{in}}\matr{D}_{\mathrm{T}}\Jmat_{\mathrm{T}} = \matr{0},\label{eqd:spieint0}
\end{align}
where the matrices in~\eqref{eqd:spieintq} were concatenated as
\begin{align}
	\Mpvmat_{\mathrm{in}} &=
	\begin{bmatrix}
		\Mpvmat^{(1)} & \cdots & \matr{0} \\
		\vdots & \ddots & \vdots \\
		\matr{0} & \cdots & \Mpvmat^{(N_{\mathrm{obj}})}
	\end{bmatrix},\label{eqd:Mpv}\\
	\Lmat_{\mathrm{in}} &=
	\begin{bmatrix}
		\Lmat^{(1)} & \cdots & \matr{0} \\
		\vdots & \ddots & \vdots \\
		\matr{0} & \cdots & \Lmat^{(N_{\mathrm{obj}})}
	\end{bmatrix},\label{eqd:L}\\
	\matr{D}_{\mathrm{T}} &=
	\begin{bmatrix}
	\frac{1}{\sigma^{(q)}}\,\matr{D}_{\mathrm{T}}^{(1)} & \cdots & \matr{0} \\
	\vdots & \ddots & \vdots \\
	\matr{0} & \cdots & \frac{1}{\sigma^{(q)}}\,\matr{D}_{\mathrm{T}}^{(N_{\mathrm{obj}})}
	\end{bmatrix},\label{eqd:DT}\\
	\Jmat_{\mathrm{T}} &=
	\begin{bmatrix}
		\Jmat_{\mathrm{T}}^{(1)} \\ \vdots \\ \Jmat_{\mathrm{T}}^{(N_{\mathrm{obj}})}
	\end{bmatrix}.\label{eqd:JT}
\end{align}
Since~$\matr{v}_{\mathrm{r}}$ has a size of~${\sum_{q}^{N_{\mathrm{obj}}}(N_{\mathrm{tri}}^{(q)}-1})$, the internal SPIE~\eqref{eqd:spieintq}, which involves testing on all mesh triangles,~\eqref{eqd:spieint0} is over-determined.
To recover a square system of equations,~\eqref{eqd:spieint0} can be left-multiplied by~$\matr{D}_{\mathrm{r}}^T$ to delete an appropriate number of equations and obtain
\begin{align}
	\matr{D}_{\mathrm{r}}^T\Mpvmat_{\mathrm{in}}\matr{D}_{\mathrm{r}}\matr{v}_{\mathrm{r}} - \frac{1}{2}\matr{D}_{\mathrm{r}}^T\matr{I}_A\matr{D}_{\mathrm{r}}\matr{v}_{\mathrm{r}} - \matr{D}_{\mathrm{r}}^T\Lmat_{\mathrm{in}}\matr{D}_{\mathrm{T}}\Jmat_{\mathrm{T}} = \matr{0},\label{eqd:spieint}
\end{align}
where the superscript~$T$ denotes taking the transpose of the associated matrix.
This operation and the choice of discretization in~\eqref{eqd:phirt0} together ensure that a full rank system matrix will eventually be obtained.

\subsection{Discrete Charge and Current Equations}\label{sec:rhokcl}

The total charge specification~\eqref{eq:Qdef} for each isolated object, and~\eqref{eq:Qdef3} for sets of objects connected to each other via ports, can all be written together in the discrete domain as
\begin{align}
	\matr{S}\ndgPhimat = \matr{Q},\label{eqd:Qcons}
\end{align}
where~$\matr{S}$ contains as many rows as the number of isolated objects plus the number of sets of objects connected to each other via ports, \emph{except} objects which are set to a given potential with respect to infinity. For example, there would be two rows for the entire setup in \figref{fig:geom3}, because there is one isolated object (\figref{fig:geom1}) plus one connected set of objects (\figref{fig:geom2}).
Each row of~$\matr{S}$ contains ones in columns corresponding to the entries of~$\ndgPhimat$ associated with that object or object set, and zeros elsewhere.
Column vector~$\matr{Q}$ contains the known total charge on each isolated object or object set, excluding objects set to a given potential with respect to infinity.
The total charge is not specified for objects connected to a given potential with respect to infinity, because those objects may draw any amount of charge necessary to maintain a potential equal to the applied potential.
For those objects, the discrete version of~\eqref{eq:phiatST} is used,
\begin{align}
	\matr{D}_0\Phimat = \matr{D}_0\matr{D}_{\mathrm{r}}\matr{v}_{\mathrm{r}} + \matr{D}_0\1\Phimat[\mathrm{a}] = \Phimat[0],\label{eqd:phiappl}
\end{align}
where~\eqref{eqd:phimat} was used, and~$\matr{D}_0$ is a sparse incidence matrix which selects entries of~$\Phimat$ where the potential is to be set.
We emphasize again that no assumptions need to be made about the distribution of the potential; only a single triangle on such an object needs to be explicitly set to a given potential.
If the object is isolated, then a constant scalar potential over the object's surface will naturally be obtained as part of the solution of the final system of equations described in~\secref{sec:finalsys}.

Finally, the discrete versions of the KVL equations~\eqref{eq:kvl1}--\eqref{eq:kvlt2} and the KCL equations~\eqref{eq:kcl1} and~\eqref{eq:kcl2} are
\begin{align}
	\matr{P}\matr{D}_{\mathrm{r}}\matr{v}_{\mathrm{r}} + \matr{P}\1\Phimat[\mathrm{a}] + \matr{R}\Jmat[\mathrm{T}] &=  \matr{V}_{\mathrm{src}},\label{eqd:kvl}\\
	\matr{C}\Jmat[\mathrm{T}] &= \matr{0},\label{eqd:kcl}
\end{align}
respectively.
In~\eqref{eqd:kvl}, matrix~$\matr{P}$ computes potential differences between terminals and its entries include~$+1$,~$-1$, and~$0$.
Matrix~$\matr{R}$ is diagonal and contains the Th\'evenin equivalent resistance associated with each port, and~$\matr{V}_{\mathrm{src}}$ contains the source voltage value at each port.
In~\eqref{eqd:kcl}, matrix~$\matr{C}$ applies the KCL for the terminals of objects, treating each object as a single node; its entries include~$+1$,~$-1$, and~$0$.
Here, we have assumed that each terminal spans a single mesh triangle.
However, a single terminal can be made to encompass multiple triangles by introducing additional equations to enforce a constant potential over the triangles spanned by the terminal~\cite{gope}.
Recall from \secref{sec:circuit:twoport} that only one of~\eqref{eq:kcl1} and~\eqref{eq:kcl2} is needed per pair of objects connected by ports.

\subsection{Final System of Equations}\label{sec:finalsys}

Finally, concatenating~\eqref{eqd:spieext},~\eqref{eqd:spieint},~\eqref{eqd:Qcons},~\eqref{eqd:phiappl},~\eqref{eqd:kvl}, and~\eqref{eqd:kcl} gives the system of equations which must be solved,
\begin{align}
	{\small%
	\setlength{\arraycolsep}{2.0pt}
	\begin{bmatrix}
		\Lmat & \Mmat\matr{D}_{\mathrm{r}} &  \1 & \matr{0} \\
		\matr{0} & \matr{D}_{\mathrm{r}}^T\Mmat_{\mathrm{in}}\matr{D}_{\mathrm{r}}& \matr{0} & -\matr{D}_{\mathrm{r}}^T\Lmat_{\mathrm{in}}\matr{D}_{\mathrm{T}} \\
		\matr{S} & \matr{0} & \matr{0} & \matr{0} \\
		\matr{0} & \matr{D}_0\matr{D}_{\mathrm{r}} & \matr{D}_0\1 & \matr{0} \\
		\matr{0} & \matr{P}\matr{D}_{\mathrm{r}} & \matr{P}\1 & \matr{R} \\
		\matr{0} & \matr{0} & \matr{0} & \matr{C}
	\end{bmatrix}
	\begin{bmatrix}
		\ndgPhimat \\ \matr{v}_{\mathrm{r}} \\ \Phimat[\mathrm{a}] \\ \Jmat_{\mathrm{T}}
	\end{bmatrix}
	=
	\begin{bmatrix}
		-\Phimat[\mathrm{im}] \\ \matr{0} \\ \matr{Q} \\ \Phimat[0] \\ \matr{V}_{\mathrm{src}} \\ \matr{0}
	\end{bmatrix}},\label{eqd:sys}
\end{align}
where
\begin{align}
	\Mmat &= \left(\Mpvmat + \frac{1}{2}\matr{I}_A\right),\label{eqd:Mmat}\\
	\Mmat_{\mathrm{in}} &= \left(\Mpvmat_{\mathrm{in}} - \frac{1}{2}\matr{I}_A\right).\label{eqd:Mmatin}
\end{align}
A key point is that the matrix~$\matr{D}_{\mathrm{r}}^T\Mmat_{\mathrm{in}}\matr{D}_{\mathrm{r}}$ has full rank and is well conditioned, unlike~$\Mmat_{\mathrm{in}}$, which contains an approximate null space associated with constant potentials.
The system matrix in~\eqref{eqd:sys} also has full rank as a result.

\section{Results}\label{sec:results}

The proposed formulation~\eqref{eqd:sys} is tested in several scenarios, including capacitance and resistance extraction.
For simplicity, a direct solver based on LU factorization~\cite{trefethen} was used for solving~\eqref{eqd:sys} in all cases, though the discretized integral operators in~\eqref{eqd:sys} are amenable to the use of acceleration techniques coupled with iterative solvers~\cite{AIMbles,pfftmain,FMAorig,enghetaFMA}.
First, we will consider canonical capacitance and resistance extraction problems where a comparison to analytical results is possible.
Then, we will provide a comparison of the proposed method to a commercial tool for more complex structures.

\subsection{Spherical Capacitor}\label{sec:results:sphcap}

\begin{figure}[t]
	\centering
	\includegraphics[width=.8\linewidth,trim={0 0 0 0},clip=true]{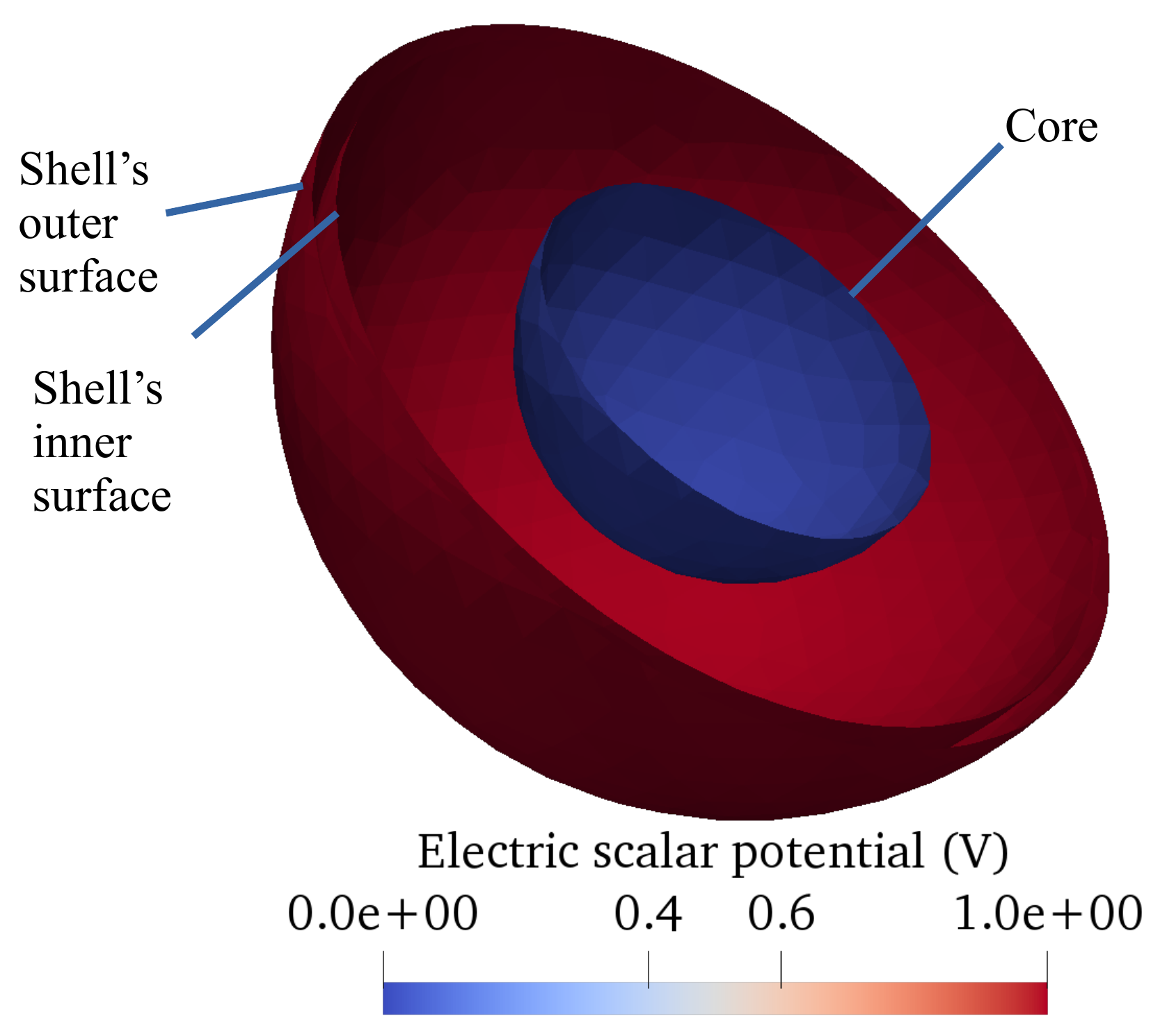}
	\caption{Cross sectional cut of the geometry and scalar potential distribution for the spherical capacitor in \secref{sec:results:sphcap}. The potential is referenced to infinity.}\label{fig:sphcap:geom}
\end{figure}

\begin{figure}[t]
	\centering
	\includegraphics[width=\linewidth]{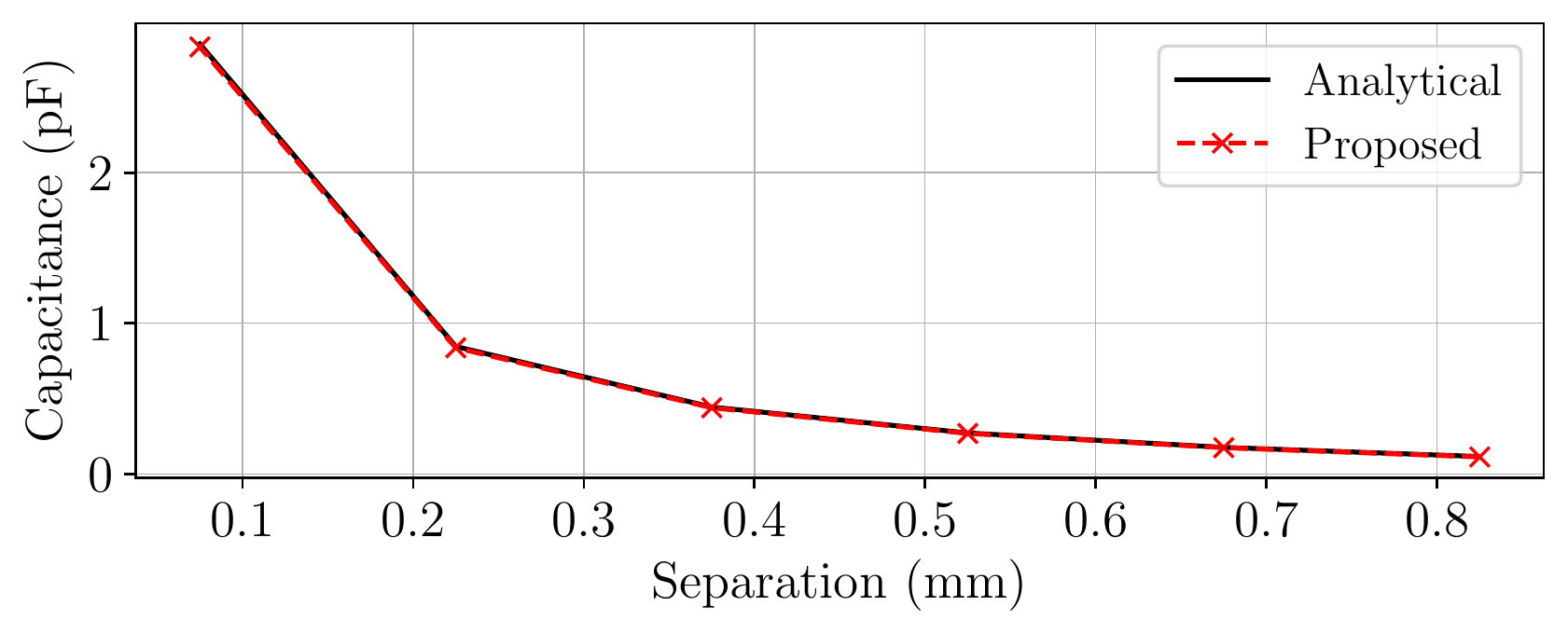}
	\caption{Capacitance of the spherical capacitor in \secref{sec:results:sphcap}.}\label{fig:sphcap:C}
\end{figure}

We consider a spherical capacitor which consists of a spherical shell of outer radius~\SI{1.5}{\milli\metre} and thickness~\SI{75}{\micro\metre}, concentrically surrounding a solid spherical core of variable radius.
A cross sectional cut in perspective is shown in \figref{fig:sphcap:geom}
Both the shell and the core are made of copper.
To compute the capacitance of the structure, a potential of~\SI{1}{\volt} is applied to a randomly-chosen triangle on the shell, and a potential of~\SI{0}{\volt} is applied to a randomly-chosen triangle on the core.
The number of mesh triangles ranged from~$3{,}400$ to~$4{,}540$ depending on the radius of the core.
System~\eqref{eqd:sys} is solved for different radii of the core, and the capacitance is computed as a post processing step by adding the elements of~$\epsilon_0\ndgPhimat$ on each object to obtain the total charge on the shell and on the core.
This allows extracting one column of the~$2\times2$ capacitance matrix of the structure, containing the self capacitance of the shell with respect to infinity and the mutual capacitance between the shell and the core.
The mutual capacitance as a function of the separation between the shell and the core is compared to the analytical result to verify the accuracy of the proposed method, as confirmed in \figref{fig:sphcap:C}.
\figref{fig:sphcap:geom} shows the electric scalar potential distribution.
A crucial point is that a constant scalar potential based on the applied potential is \emph{obtained naturally} when solving~\eqref{eqd:sys}, unlike in existing capacitance extraction formulations where a constant potential must be assumed upfront.

\subsection{Rectangular Conductor}\label{sec:results:rectcond}

\begin{figure}[t]
	\centering
	\includegraphics[width=\linewidth]{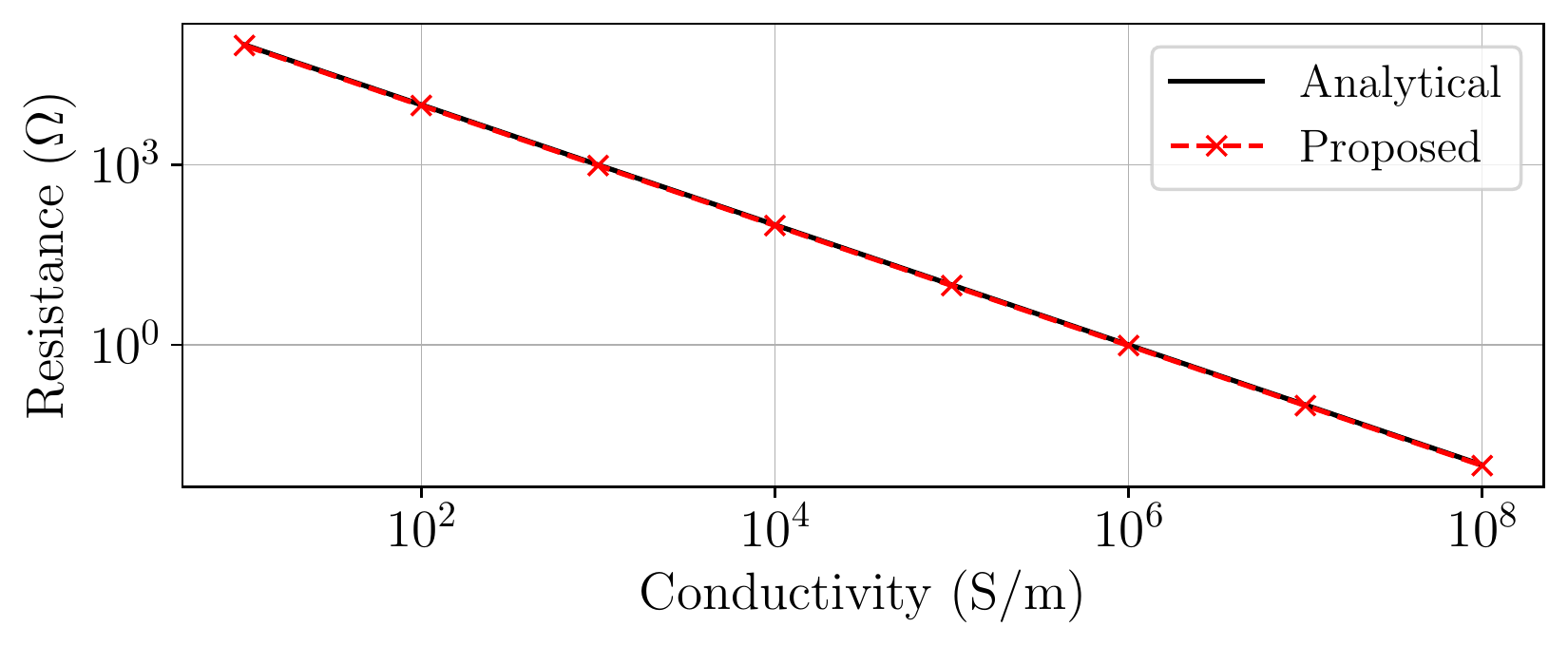}
	\caption{Resistance of the rectangular conductor in \secref{sec:results:rectcond}.}\label{fig:rectcond:R}
\end{figure}

Next, we consider a canonical resistance extraction problem involving a single rectangular prism with cross section~\SI{20}{\micro\metre}~$\times$~\SI{20}{\micro\metre} and length~\SI{0.4}{\milli\metre}, meshed with~$1{,}306$ triangles.
A port is set up as in \figref{fig:geom1} where~$\mathcal{S}_{\mathrm{T}1}$ and~$\mathcal{S}_{\mathrm{T}2}$ are defined on triangles on opposite sides of the prism along its length.
The resistance is computed easily once~\eqref{eqd:sys} is solved and~$\matr{J}_{\mathrm{T}}$ is computed.
We consider a variety of materials with varying values of conductivity and compare the resistance to the analytical formula for a conductor with a rectangular cross section,
\begin{align}
	R = \frac{l}{\sigma A},\label{eq:pouillet}
\end{align}
where~$R$,~$\sigma$,~$l$, and~$A$ are the resistance, conductivity, length, and cross section area of the rectangular conductor.
\figref{fig:rectcond:R} demonstrates the accuracy of the proposed method over nine orders of magnitude of conductivity, encompassing that of lossy dielectrics, semiconductors, and good conductors.
This indicates the generality and broad applicability of the proposed formulation.

\subsection{Part of a Capacitive Touch Sensor Panel}\label{sec:results:tsp}

\begin{figure}[t]
	\centering
	\subfloat[][]{%
		\includegraphics[width=\linewidth]{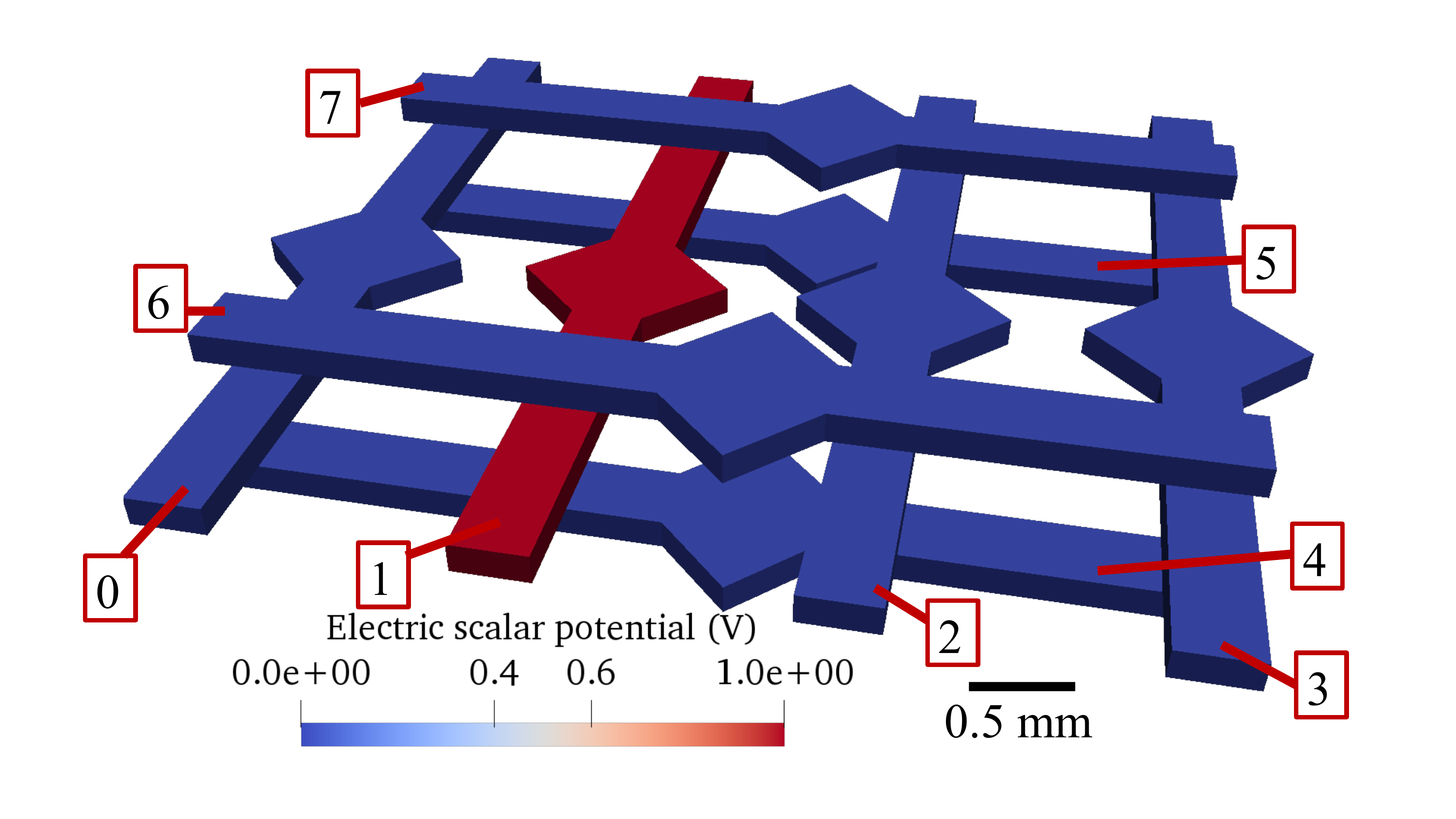}\label{fig:tspphi}}\\
	\subfloat[][]{%
		\includegraphics[width=\linewidth]{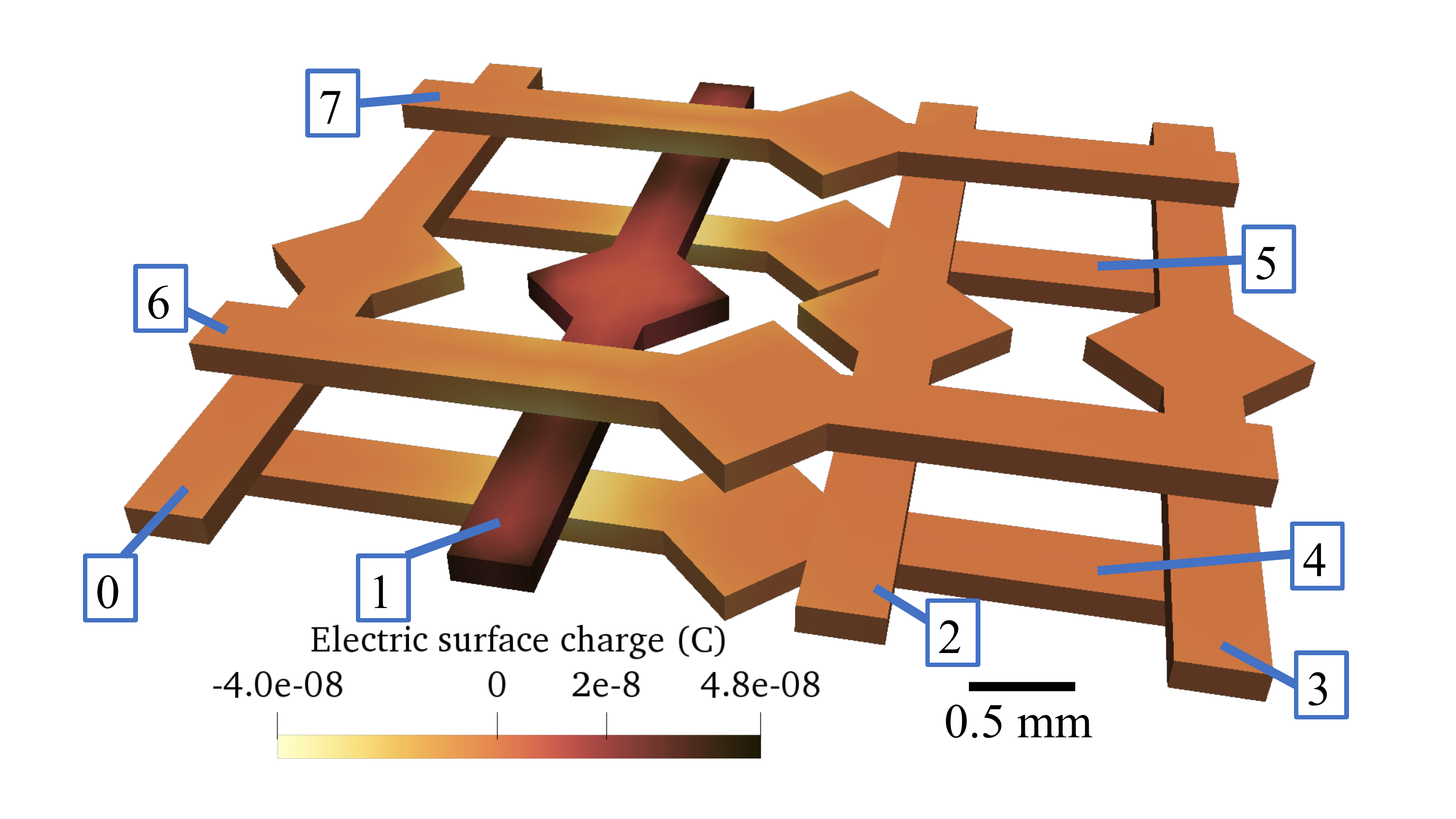}\label{fig:tsprho}}
	\caption{(a) Electric scalar potential referenced to infinity and (b) surface charge distribution for the capacitive touch sensor panel in \secref{sec:results:tsp} when a~$1\,$V potential is applied to the conductor labeled as ``1''.}\label{fig:tspgeom}
\end{figure}

\begin{figure}[t]
	\centering
	\includegraphics[width=.8\linewidth]{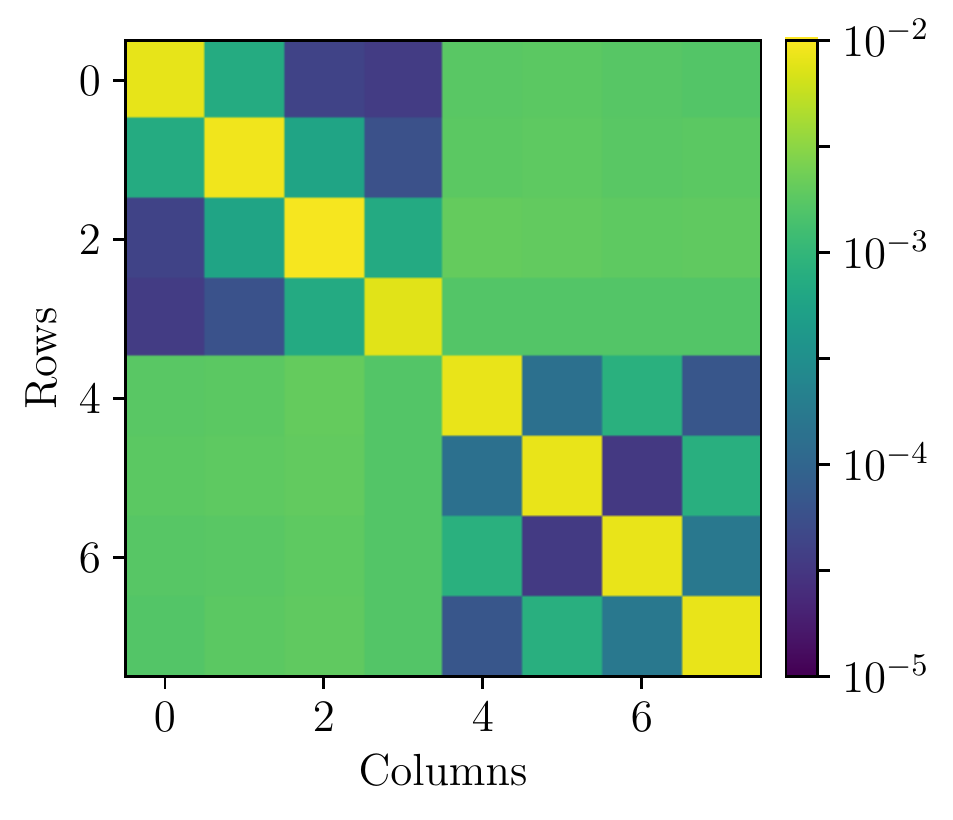}
	\caption{Relative error on a logarithmic scale for each entry of the capacitance matrix for the structure in \secref{sec:results:tsp}.}\label{fig:tsp:C}
\end{figure}

As a realistic capacitance extraction problem, we consider a structure inspired from the one considered in~\cite{tspSameer}, representing part of a flexible touch sensor panel.
The arrangement of conductors considered is shown in \figref{fig:tspgeom}, where the objects are labelled based on the order in which they will appear in the capacitance matrix.
The structure was meshed with~$3{,}108$ triangles.
On each of the eight objects, a triangle is chosen at random as a terminal to which a potential of~$1\,$V or~$0\,$V is applied.
By changing the terminal to which~$1\,$V is applied, eight separate simulations are performed to extract the entire~$8\times8$ capacitance matrix of the structure, which is then compared with results obtained from Ansys Q3D~\cite{q3d}, a commercial quasistatic solver.
\figref{fig:tspphi} shows the electric scalar potential distribution when one of the conductors is excited, and demonstrates that the resulting scalar potential is constant on each object.
Recall that no assumption of a constant scalar potential was made, unlike conventional capacitance extraction techniques.
\figref{fig:tsprho} shows the electric charge distribution on the object, and \figref{fig:tsp:C} shows the relative error in each element of the capacitance matrix on a logarithmic scale, for the proposed method compared to Ansys Q3D.
At worst, the relative error is still below~$1\,\%$, demonstrating the accuracy of the proposed technique.

\subsection{Cylindrical Via}\label{sec:results:via}

\begin{figure}[t]
	\centering
	\includegraphics[width=\linewidth]{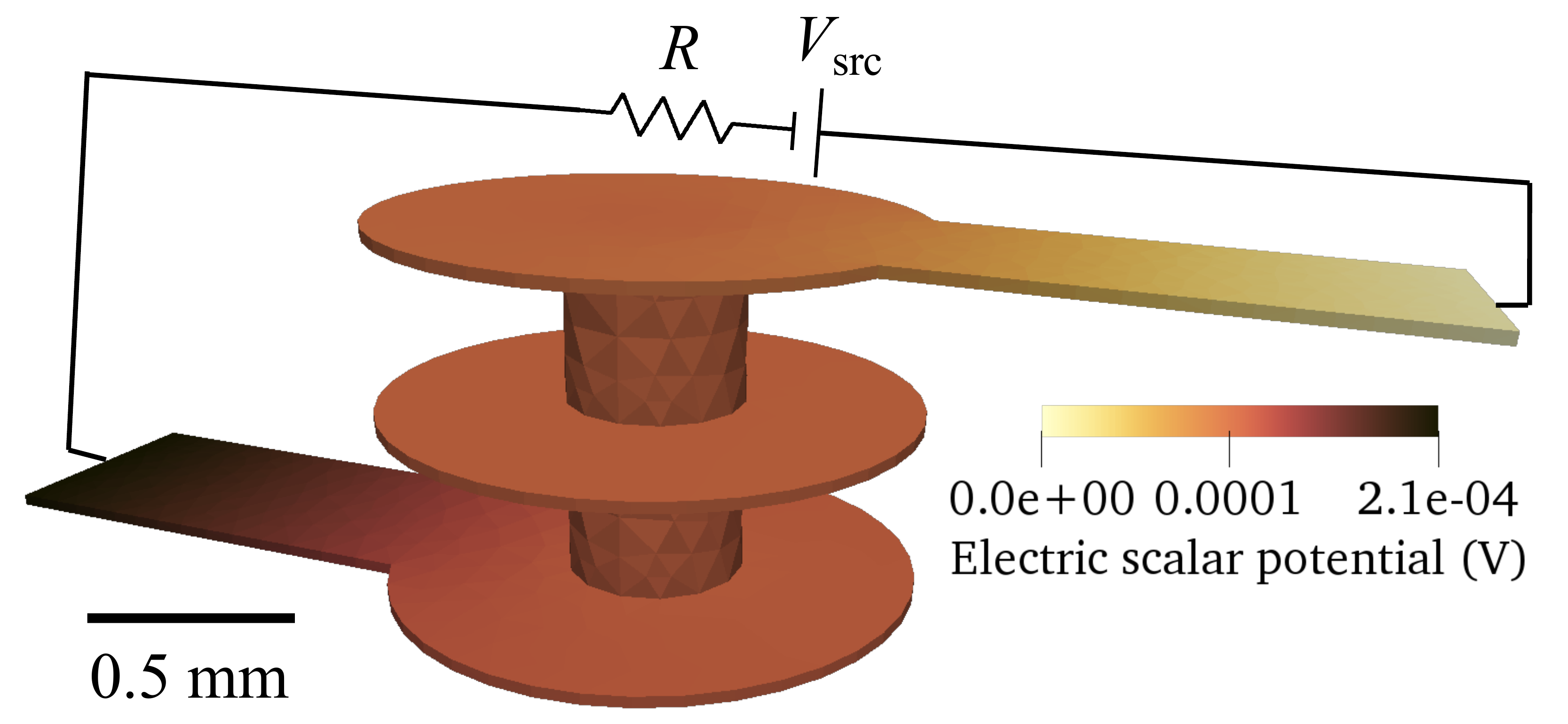}
	\caption{Geometry and scalar potential distribution (referenced to infinity) for the cylindrical via in \secref{sec:results:via} for~${V_{\mathrm{src}}=1\,}$V and~${R=50\,\Omega}$.}\label{fig:via:geom}
\end{figure}

\begin{figure}[t]
	\centering
	\includegraphics[width=\linewidth]{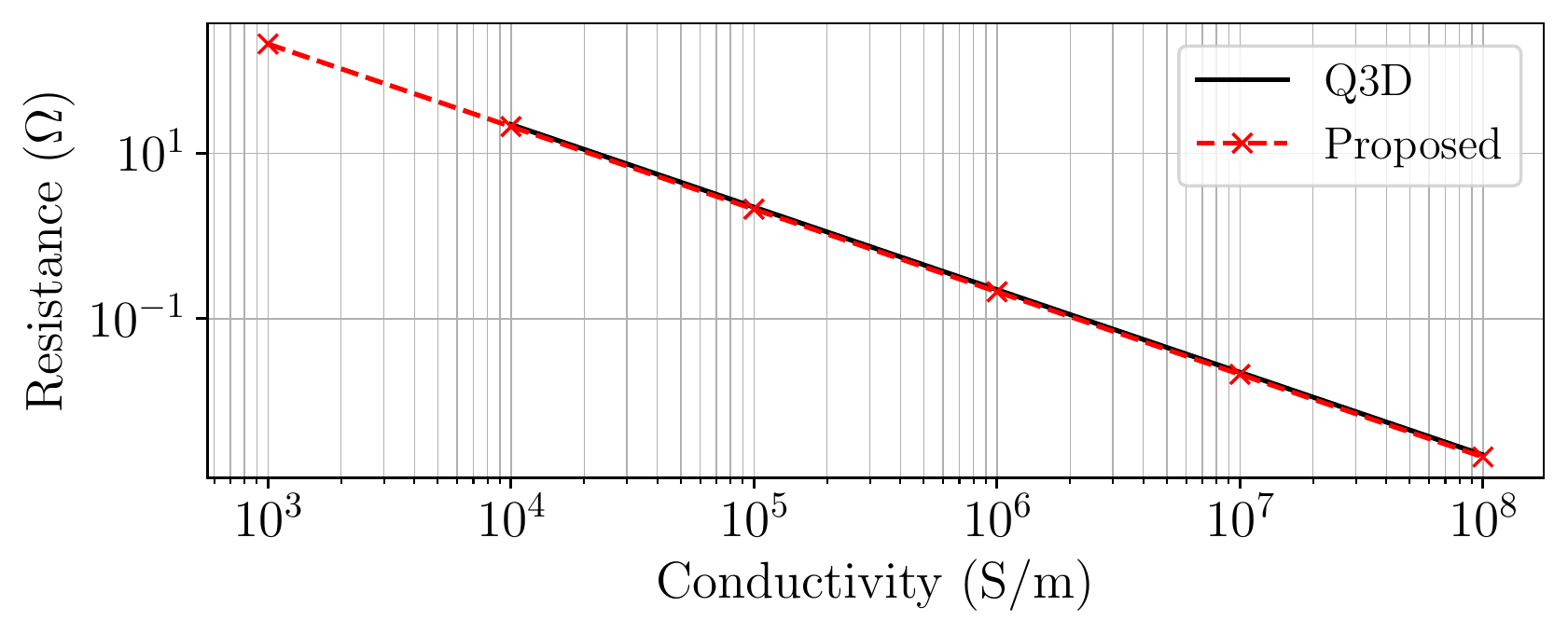}
	\caption{Resistance of the cylindrical via in \secref{sec:results:via} as a function of conductivity.}\label{fig:via:R}
\end{figure}

As a realistic resistance extraction problem, we consider a model of a cylindrical via shown in \figref{fig:via:geom}, taken from the examples provided in the commercial software Ansys Electronics Desktop~\cite{aedt}.
The structure was meshed with~$3{,}616$ triangles and excited by a~$1\,$V source in series with a~$50\,\Omega$ resistor.
The cylindrical plates have a diameter of~{\SI{1.2}{\milli\metre} and a thickness of~{\SI{25}{\micro\metre}, and they are separated, center-to-center, by a distance of~{\SI{0.4}{\milli\metre}.
The inner cylinder has a diameter of~{\SI{0.4}{\milli\metre} and a total height of~{\SI{0.8}{\milli\metre}.
The rectangular segments on either side of the via have a width of~{\SI{0.5}{\milli\metre}, a height of~{\SI{25}{\micro\metre}, and a length of~{\SI{1,1}{\milli\metre}.
The resistance of the structure is extracted for a range of conductivities and the results are compared to those obtained from Ansys Q3D~\cite{q3d}.
\figref{fig:via:geom} shows the electric scalar potential distribution for a conductivity of~$10^7\,$S/m.
\figref{fig:via:R} confirms that the proposed method can compute the DC resistance of a complex structure accurately over a wide range of conductivities spanning five orders of magnitude.
The commercial tool Q3D is geared towards highly conductive objects and therefore could not be used for conductivity values below~$10^4\,$S/m, while the proposed method remains accurate and numerically stable for both low and high conductivities.

\subsection{Part of an Interconnect Network}\label{sec:results:int}

\begin{figure}[t]
	\centering
	\includegraphics[width=\linewidth]{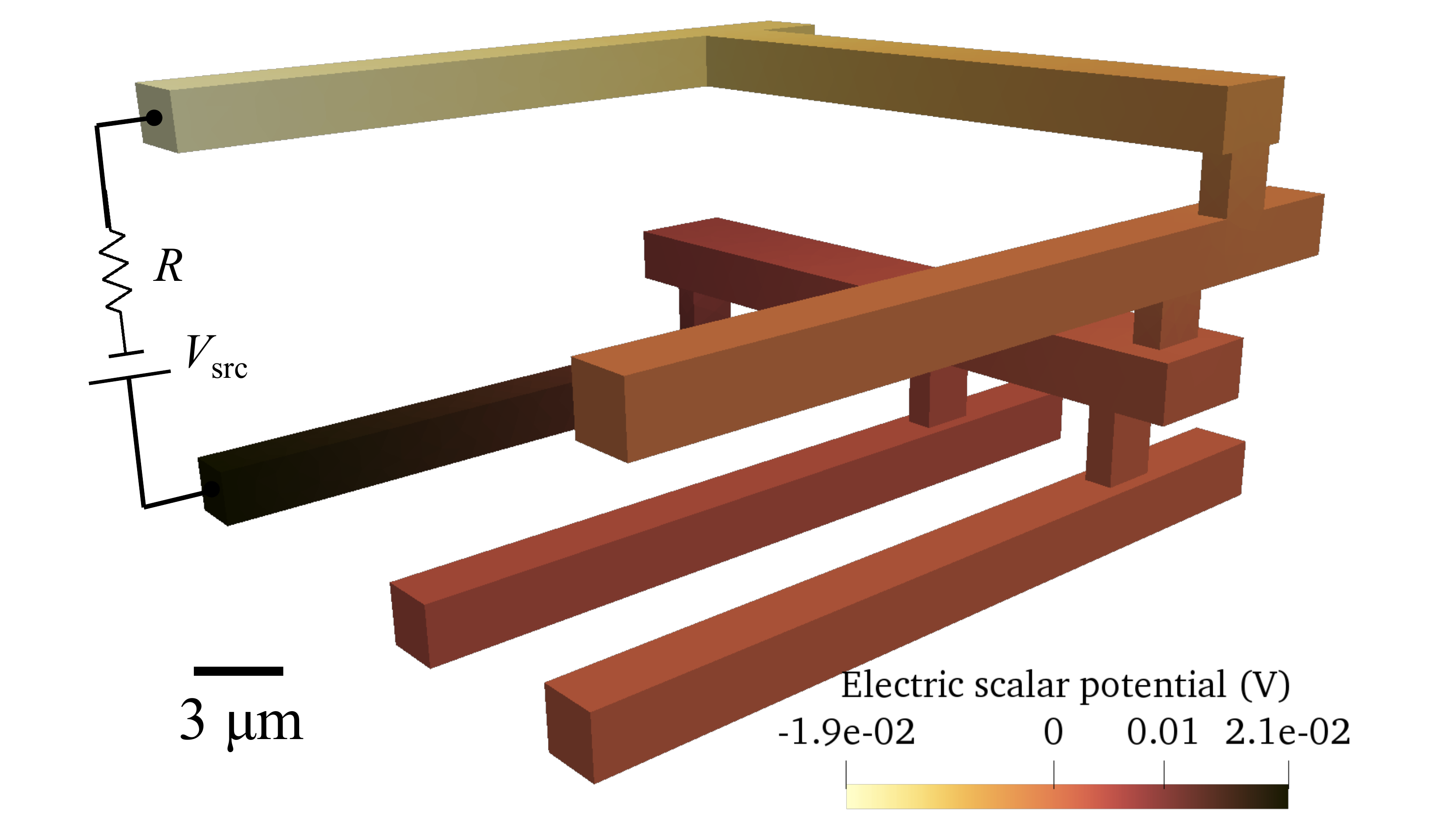}
	\caption{Geometry and scalar potential distribution for the interconnect in \secref{sec:results:int} for~${V_{\mathrm{src}}=1\,}$V and~${R=50\,\Omega}$.}\label{fig:int:geom}
\end{figure}

\begin{figure}[t]
	\centering
	\includegraphics[width=\linewidth]{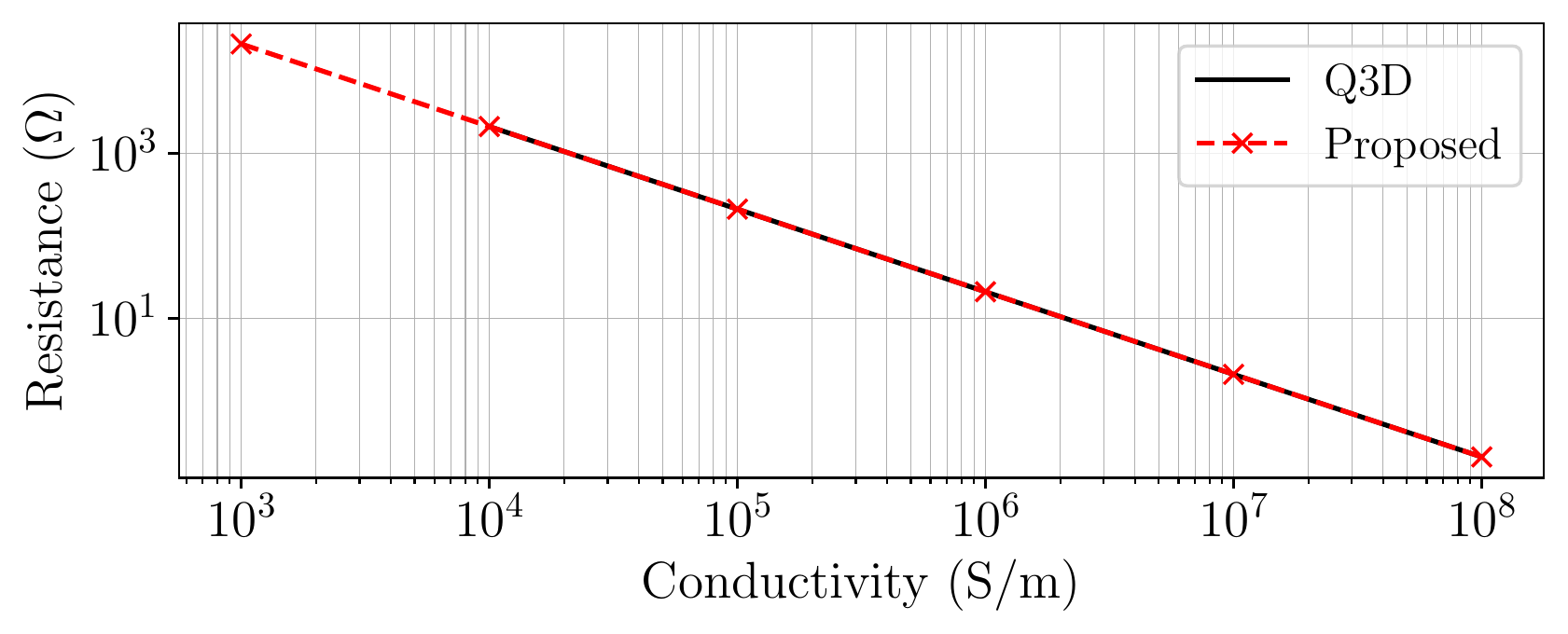}
	\caption{Resistance of the interconnect in \secref{sec:results:int} as a function of conductivity.}\label{fig:int:R}
\end{figure}

Next, we consider another resistance extraction problem involving the part of an interconnect network shown in \figref{fig:int:geom}.
The geometry of this structure was inspired by the one considered in~\cite[Chap. 5]{book:DCRC}, and the resistance is computed over a wide range of conductivities.
Each rectangular segment shown in \figref{fig:int:geom}, except the two upper-most segments, has a cross section of~{\SI{2}{\micro\metre}$\,\times\,$\SI{2}{\micro\metre}}, while the upper two segments have a width of~\SI{3}{\micro\metre} and a height of~\SI{2}{\micro\metre}.
All the rectangular segments have a length of~\SI{28}{\micro\metre}.
The vertical vias connecting the segments have a height of~\SI{2}{\micro\metre} and a cross section of~{\SI{1}{\micro\metre}$\,\times\,$\SI{2}{\micro\metre}}.
The structure was meshed with~$5{,}042$ triangles.
\figref{fig:int:geom} shows the electric scalar potential distribution for a conductivity of~$10^7\,$S/m, and \figref{fig:int:R} demonstrates that the proposed method yields accurate resistance values for a wider range of conductivity than does the commercial tool Ansys Q3D.
Again, Q3D is unable to provide a solution for a conductivity below~$10^4\,$S/m.

\subsection{Resistance and Capacitance in One Simulation}\label{sec:results:par}

\begin{figure}[t]
	\centering
	\includegraphics[width=.6\linewidth]{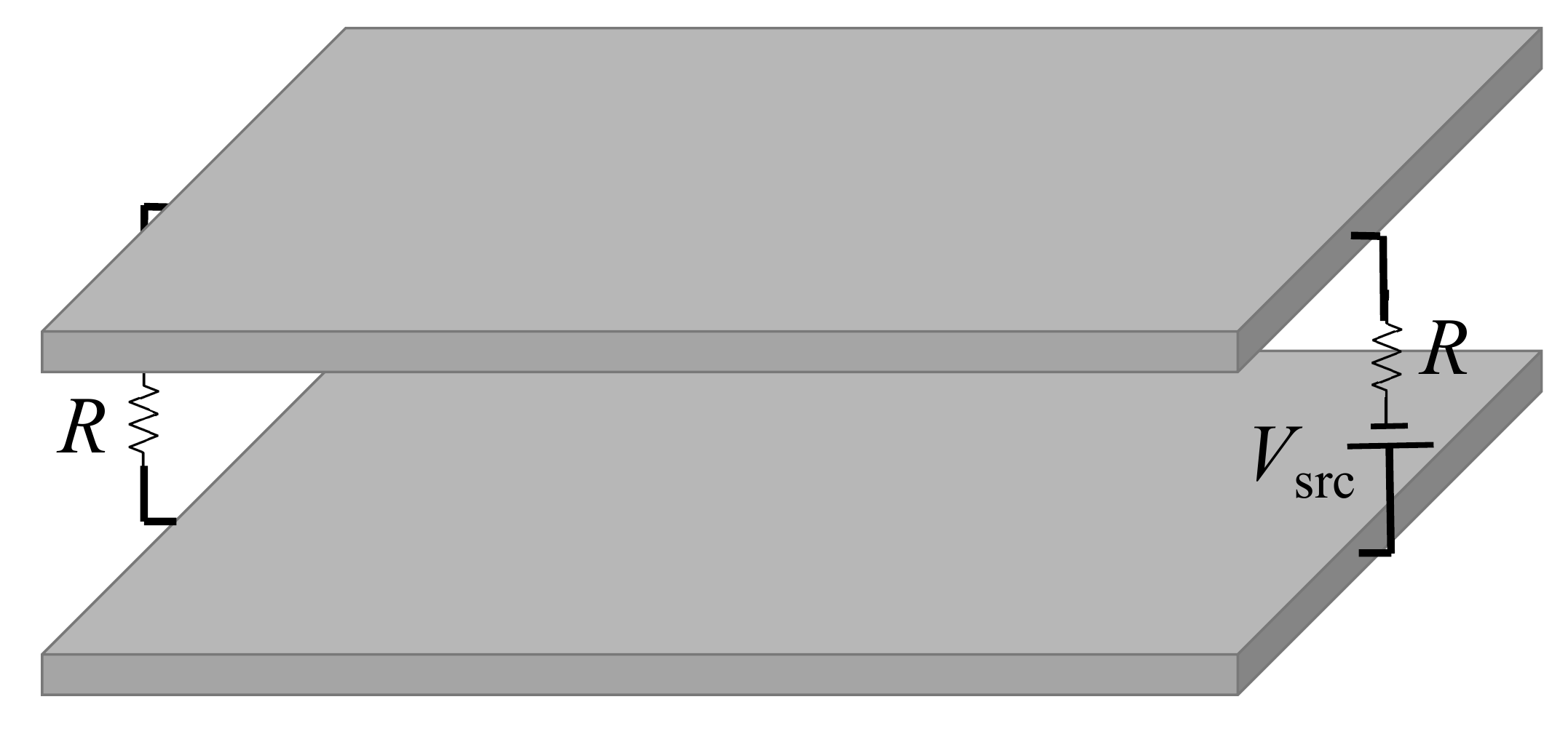}
	\caption{Geometry and excitation used for the parallel plate capacitor in \secref{sec:results:par}.}\label{fig:par:geom}
\end{figure}

We consider here an example of using the proposed method to compute both the resistance and capacitance in a single simulation.
A finite-sized parallel plate capacitor is modeled (\figref{fig:par:geom}), meshed with~${5,008}$ triangles.
Each plate is composed of copper with a conductivity of~$5.8\times 10^7\,$S/m, has a width and length of~\SI{0.5}{\milli\meter}, and has a height of~\SI{0.01}{\milli\meter}.
The plates are separated center-to-center by a distance of~\SI{0.05}{\milli\meter}.
The plates are included as part of a two-port network as in \figref{fig:par:geom}, where the resistors in the Th\'evenin circuit have a value of~$50\,\Omega$, while the voltage source supplies~$1\,$V.
Two approaches may be taken: one may attach a terminal to an arbitrary small area on each plate spanning one or a few mesh triangles, or the terminal may span an entire side edge of a plate.
The latter scenario is expected to provide a better match to the analytical resistance for a rectangular prism because the current will flow more uniformly across the plate, while the former is a more realistic setup when the measurement probe is much smaller than the width of the plates; both approaches were simulated here.

The analytical capacitance (neglecting fringing fields) expected for a canonical parallel plate capacitor with the above dimensions is~$0.0443\,$pF.
As before, adding the elements of~$\epsilon_0\ndgPhimat$ allows computing the total charge on each plate.
Knowing the average scalar potential~$\matr{\Phi}_{\mathrm{a}}$ on each plate then allows computing the capacitance of the structure, which was found to be~$0.0597\,$pF, corresponding to a relative error of~$35\%$ compared to the analytical approximation which neglects fringing fields.
When the side length of the square plates was increased to~\SI{0.8}{\milli\meter} for the same separation, the relative error in capacitance compared to the analytical value was reduced to~$24\%$; when increased to a size of~\SI{1.3}{\milli\meter}, the relative error was~$16\%$.
This trend indicates that the error is primarily due to the finite size of the plates.

Knowledge of the port currents~$\matr{J}_{\mathrm{T}}$ and the space-dependent voltage distribution~$\matr{v}_{\mathrm{r}}$ allows computing the resistance of each plate with simple circuit analysis for the setup shown in \figref{fig:geom2}.
The analytical resistance of a rectangular plate of the chosen dimensions is~$1.724\,\text{m}\Omega$.
The resistance of the plates when excited via large terminals spanning the entire side edges of the plates was found to be~$1.525\,\text{m}\Omega$.
When excited by a small terminal spanning only a few mesh triangles each, the resistance was found to be~$2.556\,\text{m}\Omega$.
As expected, the resistance in the large-terminal case matches the analytical value more closely because the current flows more uniformly across the plates, while that in the small-terminal case is significantly higher because the current flow is no longer uniform.
This example demonstrates the unifying property of the proposed formulation: with existing BEM approaches, two different formulations would be needed to compute the capacitance and resistance of this structure.
 
\subsection{General Structure with Multiple Excitations}\label{sec:results:gen}

\begin{figure}[t]
	\centering
	\includegraphics[width=\linewidth]{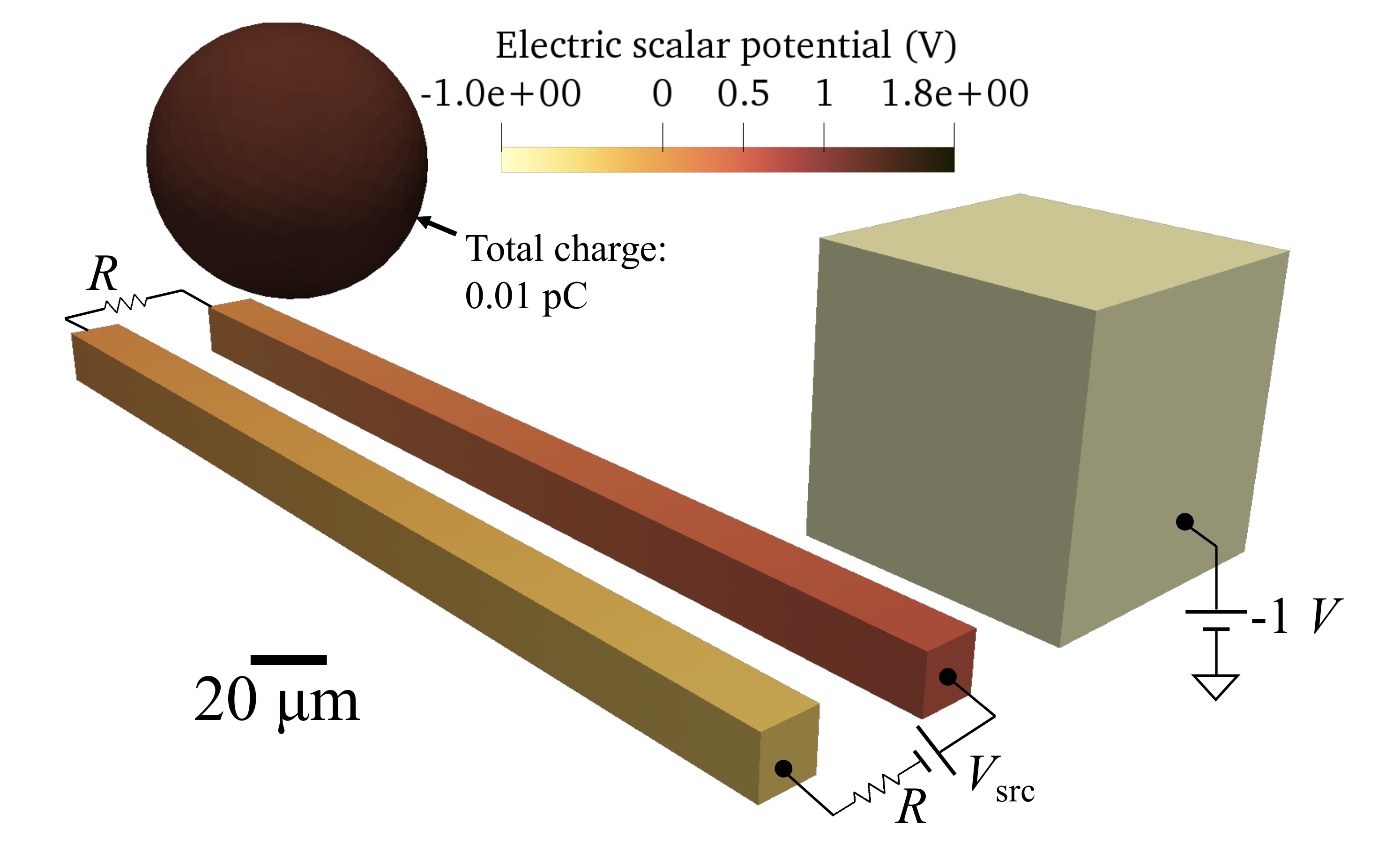}
	\caption{Geometry and scalar potential distribution for the structure in \secref{sec:results:gen} referenced to infinity.}\label{fig:gen:geom}
\end{figure}

Here, we consider a structure containing a combination of the various types of excitation supported by the proposed formulation, to demonstrate the generality of the method.
The structure is shown in \figref{fig:gen:geom}, and consists of a differential pair of rectangular conductors (each with dimensions~{\SI{20}{\micro\metre}$\,\times\,${\SI{20}{\micro\metre}$\,\times\,${\SI{400}{\micro\metre}) connected in a closed circuit, a cube (with side length~{\SI{100}{\micro\metre}) with a given potential applied to a randomly-chosen triangle on the cube, and a sphere (with radius~\SI{50}{\micro\metre}) on which the total charge is specified.
Situations involving spheres with the total charge specified arise in the modeling of molecular interactions and protein folding~\cite{spiemol01,spiemol05,spiemol06}.
All objects have a conductivity of~$10\,$S/m, and a mesh with~${5,722}$ triangles was used.
On the cube, a potential of~$-1\,$V is applied, while the total charge on the sphere is set to~$0.01\,$pC.
As expected, the resulting distribution of scalar potential on the cube and sphere are constant, as shown in \figref{fig:gen:geom}, while the potential across the connected pair of conductors varies in accordance with the current flowing in the circuit.
Analytically, this implies a surface potential of~$1.798\,$V when proximity effects are ignored, which is in excellent agreement with the value of~$1.8\,$V observed in \figref{fig:gen:geom}.
From analyzing the computed port currents~$\matr{J}_{\mathrm{T}}$ and the port potentials, the resistance of each conductor of the differential pair is obtained as~$97.45\,$k$\mathrm{\Omega}$.
This deviates by approximately~$2.5\,\%$ from the analytical value of~$100\,$k$\mathrm{\Omega}$ predicted by Pouillet's law~\eqref{eq:pouillet} for a rectangular prism, which does not take into account the proximity of the cube to the sphere.

Overall, the examples considered in this section demonstrate the flexibility and generality of the proposed formulation~\eqref{eqd:sys}; it unifies various functionalities and in due course could be useful in a variety of scenarios which do not meet the assumptions of existing formulations.

\section{Conclusion}\label{sec:conclusion}

A boundary element formulation based on the electric scalar potential was proposed for the electrostatic analysis of structures composed of arbitrary conductive objects.
An intuitive and rigorous mathematical treatment was provided to handle the null space associated with operators which model the region internal to each conductive object, to yield a system of equations that has full rank.
Since no application-specific assumptions are made on the scalar potential or on the charge distribution, the proposed formulation is extremely general and may be useful in a variety of scenarios ranging from resistance and capacitance extraction to the modeling of molecular interactions.
The proposed method is simple to implement involving standard boundary element operators, and is amenable to the use of standard acceleration algorithms to model large problems.
Several numerical examples were considered to demonstrate the accuracy and generality of the proposed formulation over several orders of magnitude of material conductivity.



%

%

%
%

\ifCLASSOPTIONcaptionsoff
  \newpage
\fi



\bibliographystyle{IEEEtran}
\bibliography{./IEEEabrv,./bibliography}
\end{document}